\documentclass[a4paper,11pt]{article}
\pdfoutput=1 

\usepackage{jheppub} 

\usepackage[T1]{fontenc} 
\usepackage{graphicx}
\usepackage{amsmath, amsfonts}

\usepackage{color}

\usepackage[normalem]{ulem}  

\newcommand{\non}{\nonumber\\}

\newcommand{\be}{\begin{equation}}
\newcommand{\ee}{\end{equation}}
\newcommand{\bea}{\begin{eqnarray}}
\newcommand{\eea}{\end{eqnarray}}
\newcommand{\ba}[1]{\begin{array}{#1}}
\newcommand{\ea}{\end{array}}

\newcommand{\Tr}{{\rm Tr}}

\title{Layers of deformed instantons in holographic baryonic matter}

\author[a]{Florian Preis}\author[b]{and Andreas Schmitt}
 
\affiliation[a]{Institut f\"{u}r Theoretische Physik, Technische Universit\"{a}t Wien, 1040 Vienna, Austria} 
\affiliation[b]{Mathematical Sciences and STAG Research Centre, University of Southampton, Southampton SO17 1BJ, United Kingdom}

\emailAdd{fpreis@hep.itp.tuwien.ac.at}
\emailAdd{a.schmitt@soton.ac.uk}

\abstract{We discuss homogeneous baryonic matter in the decompactified limit of the Sakai-Sugimoto model, improving existing approximations based on flat-space instantons.
We allow for an anisotropic deformation of the instantons in the holographic and spatial directions and for a density-dependent distribution of arbitrarily many instanton layers in the bulk. 
Within our approximation, the baryon onset turns out to be a second-order phase transition, at odds with nature, 
and there is no transition to quark matter at high densities, at odds with expectations from QCD. 
This changes when we impose certain constraints on the shape of single instantons, 
motivated by known features of holographic baryons in the vacuum. Then, a first-order baryon onset and chiral restoration at high density are possible, and
at sufficiently large densities two instanton layers are formed dynamically. Our results are a further step towards describing realistic, strongly interacting matter over a large density regime within a single model, desirable for studies of compact stars.}


\begin{document} 
\maketitle
\flushbottom

\section{Introduction}
\label{sec:intro}

Cold and dense matter in the interior of compact stars is strongly interacting, governed by Quantum Chromodynamics (QCD). Its phases and properties are poorly known because
it is much denser than ordinary nuclei on earth, but not asymptotically dense and thus not quantitatively accessible with weak-coupling methods.
The gauge/string duality \cite{Maldacena:1997re,Gubser:1998bc,Witten:1998qj} is a powerful tool to study strongly interacting matter and has proven to be very useful to get insight into hot QCD matter
at low baryon densities produced in heavy-ion collisions \cite{casalderrey2014gauge}. It is thus natural to ask whether we can use it to learn something about cold and dense matter too
\cite{Ghoroku:2013gja,Kim:2014pva,Hoyos:2016zke}. 
Dense QCD is expected to have a very rich phase structure, including color-superconducting quark matter \cite{Alford:2007xm}, and at present there are no holographic approaches that can
be expected to predict reliably any details of this phase structure. 
Here we are asking a more modest question, which nevertheless may turn out to be valuable for the study of compact stars. We
are asking whether the Sakai-Sugimoto model \cite{Witten:1998zw,Sakai:2004cn,Sakai:2005yt}, a certain realization of the gauge/string duality that comes as close to QCD as currently possible,
can be used to understand at least the gross thermodynamic properties of dense nuclear and quark matter and possibly the transition between them,
ignoring all complications such as Cooper pairing of nucleons or quarks. The main point of our current effort, started in Ref.\ \cite{Li:2015uea}, is to first find a feasible 
approximation that gets the basic properties of dense matter right, and then, in future studies, to apply this approximation to the physics of compact stars. In particular,
we are interested in the onset of nuclear matter, which must show a discontinuity in the baryon density due to the finite binding energy, and in the transition to quark matter,
which is expected to happen at high densities and which
is needed to investigate hybrid stars, i.e., compact stars containing quark matter in the core, surrounded by nuclear matter. Although the Sakai-Sugimoto model is a top-down approach,
our study should not be understood as a first-principle calculation because we apply various approximations and simplifications.  
We rather aim at a model description of dense matter, which has some
advantages over many of the field-theoretical models used in the same context: we employ a genuine strong-coupling formalism, we can account for nuclear {\it and} quark matter in a single
model, and the model has very few parameters (3 in the version we consider: the 't Hooft coupling $\lambda$, the Kaluza-Klein mass $M_{\rm KK}$, and the asymptotic separation of the D8- and $\overline{\rm D8}$-branes 
$L$).

Baryons in the Sakai-Sugimoto model are introduced as D4-branes wrapped around the 4-sphere of the background geometry, following the general concept of baryons in the
gauge/string duality \cite{Witten:1998xy,Gross:1998gk}. Here, this is equivalent to gauge field configurations with nonzero topological charge on the connected flavor branes of the
model \cite{Sakai:2004cn}, and various properties of baryons in the vacuum have been studied within this approach \cite{Sakai:2004cn,Sakai:2005yt,Hata:2007mb,Seki:2008mu,Cherman:2011ve}.
Baryonic matter at nonzero density and temperature was first considered in a pointlike approximation of the instantons on the flavor branes \cite{Bergman:2007wp}, and this approach was improved
and complemented by a number of studies \cite{Rozali:2007rx,Kim:2007vd,Rho:2009ym,Kaplunovsky:2010eh,Preis:2011sp,Kaplunovsky:2012gb,deBoer:2012ij,Ghoroku:2012am,Bolognesi:2013nja,Rozali:2013fna,Bolognesi:2013jba,Bolognesi:2014dja,Kaplunovsky:2015zsa,Li:2015uea}. (For studies of baryonic matter 
using a different holographic approach, based on a D3-D7 setup, see for instance Refs.\ \cite{Gwak:2012ht,Evans:2011eu}.) The idea of the present paper is to improve the instanton gas approach, introduced in Ref.\ \cite{Ghoroku:2012am} and further developed in Ref.\ \cite{Li:2015uea}. More specifically, it is known that away from the $\lambda=\infty$ limit
the Sakai-Sugimoto instantons are anisotropic in the sense that
they break the SO(4) symmetry of rotations in the space of the holographic coordinate and the three spatial dimensions \cite{Rozali:2013fna,Bolognesi:2013jba}.
We shall account for this anisotropy by introducing a ``deformation parameter'' into the standard flat-space instanton solution. Furthermore,
it has been argued that the repulsion between the instantons makes them spread out in the holographic direction \cite{Rozali:2007rx}, which is realized for instance in crystalline structures in the
confined phase of the model \cite{Kaplunovsky:2012gb,Kaplunovsky:2015zsa}. We introduce this repulsive effect by allowing for an arbitrary number of instanton layers in the bulk and determine this number and the
distance between the layers dynamically as a function of the baryon chemical potential.

Unless backreactions of the flavor branes on the background geometry are taken into account, cold matter in the Sakai-Sugimoto model does not deconfine, which is in accordance with
expectations from large-$N_c$ QCD \cite{McLerran:2007qj}. In order to allow for a transition between nuclear and quark matter at low temperatures, we work in the ``decompactified'' limit of the model:
if the separation of the flavor branes $L$ is sufficiently small, the deconfined geometry of the model has a chirally symmetric and a chirally broken phase. Therefore,
we are able to include the transition from nuclear matter to quark matter without the complications of backreacting flavor branes. Varying $L$ from its maximum value, as used in the
original works \cite{Sakai:2004cn,Sakai:2005yt}, to very small values is best understood as changing the dual field theory: the limit of maximal $L$, i.e., an antipodal separation of the flavor branes
in the space of the compactified extra dimension of the model, is related to large-$N_c$ QCD; the limit of very small $L$, on the other hand, corresponds to a field theory
comparable to a Nambu-Jona Lasinio (NJL) model \cite{Antonyan:2006vw,Davis:2007ka,Preis:2012fh}, and its rich phase structure in the deconfined geometry is possibly closer to real-world QCD than
the antipodal limit, at least with respect to the chiral phase transition.  

The paper is organized as follows. In Sec.\ \ref{sec:setup} we explain our ansatz and derive the free energy and its stationarity equations. This is done 
by first discussing the Dirac-Born-Infeld action in Sec.\ \ref{sec:DBI}, including a very general form of the non-abelian field strengths in Sec.\ \ref{sec:general}, 
our specific ansatz for the anisotropic instantons in Secs.\ \ref{sec:anisotropic}, the approximations for our many-instanton system in Sec.\ \ref{sec:spatial}, and the symmetrized trace prescription 
in Sec.\ \ref{sec:str}. Then, in Sec.\ \ref{sec:CS}, we add the Chern-Simons contribution to obtain the full Lagrangian, and in Sec.\ \ref{sec:mini} we explain how we solve the system, including the minimization 
of the free energy. Sec.\ \ref{sec:results} is devoted to the numerical results and is divided into two subsections: in Sec.\ \ref{sec:tseytlin} we minimize the free energy with respect to all parameters of the ansatz, while
in Sec.\ \ref{sec:fix} we impose certain constraints on the shape of the single instantons, increasing the number of free parameters of our model to 5. We give our conclusions in Sec.\ \ref{sec:summary}.

\section{Setup}
\label{sec:setup}

The general setup follows numerous other works in the Sakai-Sugimoto model, and for all details and foundations of the model we refer the reader to the original works \cite{Sakai:2004cn,Sakai:2005yt} or reviews
\cite{Peeters:2007ab,Gubser:2009md,Rebhan:2014rxa}; the notation we are using is consistent with Ref.\ \cite{Li:2015uea}. 
Our starting point is the action for the gauge fields on the flavor branes, which consists of a Dirac-Born-Infeld (DBI) and a Chern-Simons (CS) part,
\be
S = S_{\rm DBI} + S_{\rm CS} \, . 
\ee
We now discuss these two contributions separately.

\subsection{Dirac-Born-Infeld action}
\label{sec:DBI}

\begin{figure}[t]
\centering 
\includegraphics[width=0.95\textwidth]{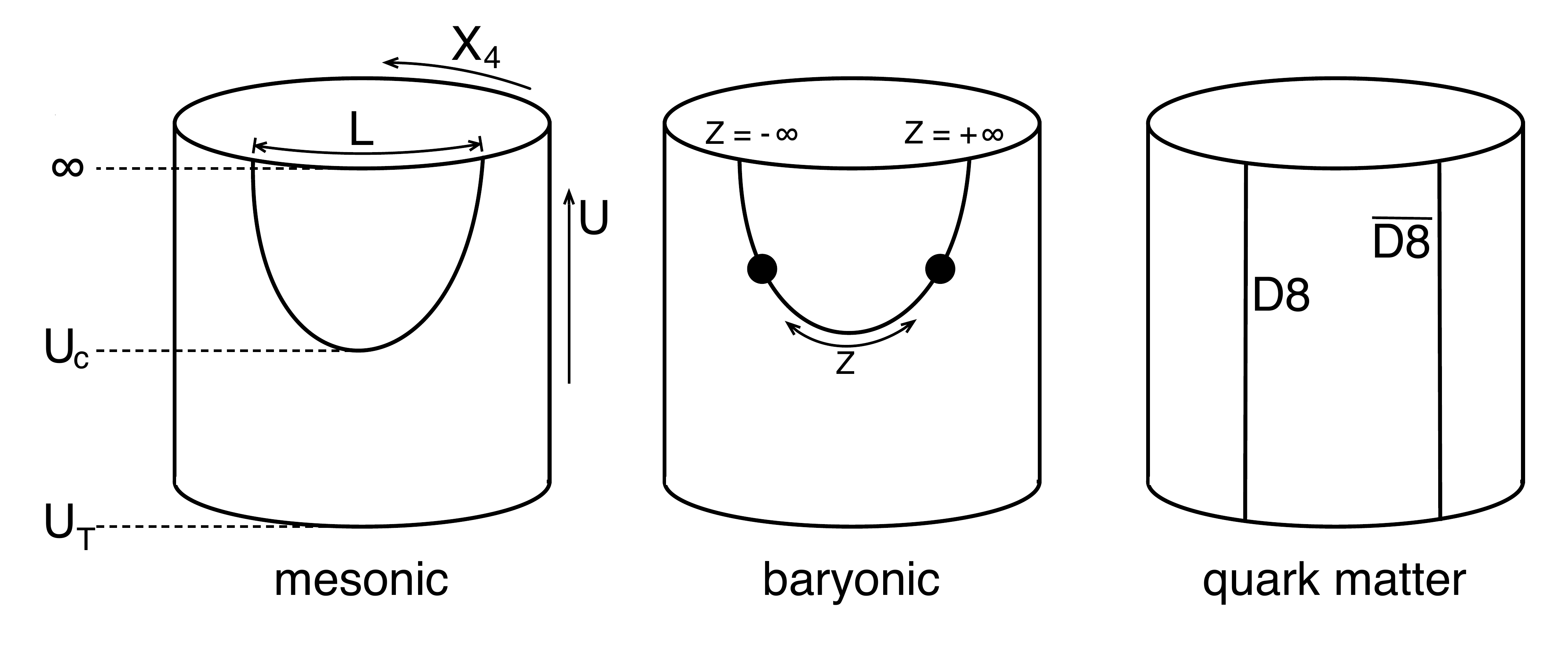}
\caption{\label{fig:cylinders} Illustration of the three phases whose free energies are compared in this paper. It shows the cylinder-shaped subspace of the deconfined geometry, spanned by the compact 
extra dimension $X_4$ (with radius $M_{\rm KK}^{-1}$) and the holographic coordinate $U$, and the D8- and $\overline{\rm D8}$-branes, which can either be connected (chiral symmetry spontaneously broken, left and middle figure) or disconnected 
(chiral symmetry restored, right figure). 
They are asymptotically separated by a distance $L$, and in the chirally broken phases their (density-dependent) embedding, including the location of the tip $U_c$, has to be determined dynamically (we assume no backreaction on the background geometry). 
Baryon number in the chirally broken phase is introduced through instantons on the flavor branes, here
symbolized by two circles. Our ansatz allows for an arbitrary number of instanton layers $N_z$ in the bulk (see Sec.\ \ref{sec:spatial}), but we shall find that more than two are never energetically preferred. 
We assume that $L\ll \pi/M_{\rm KK}$, which is the "decompactified" limit of the model, where the deconfined geometry extends down to arbitrarily small temperatures (apart from using this fact, which allows us to set 
the temperature to zero, $U_T=0$, we never make any assumptions about $L$ and $M_{\rm KK}$ in our calculation). 
For the discussion of the instantons we sometimes switch to an alternative holographic coordinate $Z$ along the connected branes, as indicated in the middle figure.}
\end{figure}

The DBI part is given by 
\be \label{SDBI}
S_{\rm DBI} = 2T_8 V_4\int_0^{1/T} d\tau \int d^3X \int_{U_c}^\infty dU \, e^{-\Phi}\,{\rm str}\sqrt{{\rm det}(g+2\pi\alpha' {\cal F})} \, . 
\ee
Here, the integral is taken over imaginary time $\tau$ with the temperature $T$, over position space $\vec{X}=(X_1,X_2,X_3)$,
and over the holographic coordinate $U\in [U_c,\infty].$ In this section, we discuss the chirally broken geometry, where the D8- and $\overline{\rm D8}$-branes are connected, with $U_c$ being the location of the 
tip of the connected branes. The model contains a compactified direction $X_4$, whose radius is expressed in terms of the inverse Kaluza-Klein mass $M_{\rm KK}$, 
$X_4 \equiv X_4 + 2\pi/M_{\rm KK}$, and the embedding of the flavor branes
in the background geometry is given by $X_4(U)$. This function has to be determined dynamically and is subject to the boundary condition $X_4(U\rightarrow\infty)=\pm L/2$, where $L$ is the asymptotic separation of the D8- and $\overline{\rm D8}$-branes. The chirally broken geometry accommodates the baryonic phase,
discussed in this section, and the mesonic phase, which is well known
and whose free energy we shall later need and simply quote from the literature. In the chirally restored phase the flavor branes are straight, $X_4(U)=\pm L/2$, and disconnected, and
again we will quote the corresponding free energy later and use it in the energy comparison with the baryonic phase. The geometry of the three different phases is shown schematically in Fig.\ \ref{fig:cylinders}.
The dilaton field is $e^{\Phi}= g_s(U/R)^{3/4}$, where $R$ is the curvature radius and $g_s$ the string coupling. Moreover, $\alpha'=\ell_s^2$ with the string length $\ell_s$, $T_8 = 1/[(2\pi)^8\ell_s^9]$ is the D8-brane tension, $V_4=8\pi^2/3$ is the volume of the 4-sphere, and ``str'' denotes the symmetrized trace (we shall discuss below the procedure that we
follow to evaluate this trace). 
We work in the deconfined geometry, whose induced metric on the flavor branes $g$ is given by  
\bea
ds_{\rm D8}^2 &=& \left(\frac{U}{R}\right)^{3/2}[f_T(U)d\tau^2+\delta_{ij}dX^idX^j] \non[2ex]
&&+\left(\frac{R}{U}\right)^{3/2}\left\{\left[\frac{1}{f_T(U)}+\left(\frac{U}{R}\right)^3
(\partial_UX_4)^2\right] dU^2 +U^2d\Omega_4^2\right\} \, ,
\eea
where $i=1,2,3$, $d\Omega_4^2$ is the metric of the 4-sphere, and we have abbreviated 
\be
f_T(U) \equiv  1-\frac{U_T^3}{U^3} \,  ,
\ee
where $U_T$ is related to temperature $T$ and curvature radius $R$ via
\be
T = \frac{3}{4\pi}\frac{U_T^{1/2}}{R^{3/2}} \, . 
\ee
In our final results we shall restrict ourselves to $T=0$. Strictly speaking, the preferred geometry at zero temperature is the confined one. However, in the decompactified limit $L\ll \pi/M_{\rm KK}$, 
the critical temperature for deconfinement is much smaller than the critical temperature for chiral restoration (at zero chemical potential). One may think of letting $M_{\rm KK}\to 0$ while keeping $L$ fixed;
this renders the region of the confined geometry in the phase diagram arbitrarily small and justifies our zero-temperature approximation. 
In the decompactified limit, the structure of the phase diagram (without baryonic matter) is similar to what is obtained in an NJL model,
where there is no confinement either. Two differences to NJL are that our formalism allows for a well-defined way to implement baryons
(which are rarely included in NJL studies, although it is possible \cite{Alkofer:1994ph}) and that in the NJL model it is easy to include nonzero current quark masses (which is difficult in the
Sakai-Sugimoto model, although it is possible \cite{Bergman:2007pm,Dhar:2007bz,Hashimoto:2008sr,Brunner:2015oga}). As a consequence, in our calculation, the chiral phase transition is always a phase transition 
in the strict sense (in fact, it turns out to be always a first-order phase transition), while in the NJL model with quark masses (and in nature) this transition is allowed to be a 
continuous crossover because chiral symmetry is not an exact symmetry.

We work with two flavors, $N_f=2$, and express the $U(2)$ field strengths in terms of the gauge fields ${\cal A}_\mu$, $\mu=0,1,2,3,U$, in which we separate the abelian from the non-abelian part,
\bea
{\cal A}_\mu &=& \hat{A}_\mu + A_\mu \, , \qquad A_\mu = A_\mu^a\sigma_a \, , 
\eea
with the Pauli matrices $\sigma_a$, normalized such that $[\sigma_a,\sigma_b]=2i\epsilon_{abc}\sigma_c$. Consequently, with 
${\cal F}_{\mu\nu} = \partial_\mu {\cal A}_\nu - \partial_\nu {\cal A}_\mu + i[{\cal A}_\mu,{\cal A}_\nu]$ we have
\bea
{\cal F}_{\mu\nu} &=& \hat{F}_{\mu\nu} + F_{\mu\nu} \, , \qquad F_{\mu\nu}= F_{\mu\nu}^a \sigma_a \, , 
\eea
with $\hat{F}_{\mu\nu} =  \partial_\mu \hat{A}_\nu - \partial_\nu \hat{A}_\mu$, and $F_{\mu\nu}^a =   \partial_\mu A_\nu^a - \partial_\nu A_\mu^a -2\epsilon_{abc}A_\mu^b A_\nu^c$. 
In our ansatz the only nonzero abelian field strength is $\hat{F}_{0U}$, where the quark chemical potential will be included as the boundary value of $\hat{A}_0$, and the baryons are implemented through
the non-abelian field strengths $F_{iU}$, $F_{ij}$. All other field strengths are set to zero. 

The trace over the square root in the DBI action is not uniquely defined in the non-abelian case, and thus we need to follow a certain prescription. Here we follow
Tseytlin \cite{Tseytlin:1999dj,Kaplunovsky:2010eh}: we first compute the determinant over space-time indices
as if the field strengths were numbers,
\bea \label{DBIdeconf}
{\rm det}(g+2\pi\alpha'{\cal F}) &=& U^8\left(\frac{R}{U}\right)^{3/2}\left\{f_T(2\pi\alpha')^2F_{iU}^2
+\left[1+\left(\frac{U}{R}\right)^3 f_T (\partial_U X_4)^2 +(2\pi\alpha')^2\hat{F}_{0U}^2\right]\right. \non[2ex]
&&\left. \times\left[1+\left(\frac{R}{U}\right)^3\frac{(2\pi\alpha')^2F_{ij}^2}{2}\right] + \left(\frac{R}{U}\right)^3\frac{f_T(2\pi\alpha')^4(F_{ij}F_{kU}\epsilon_{ijk})^2}{4}\right\} \, .
\eea
This expression factorizes if $(F_{ij}F_{kU}\epsilon_{ijk})^2=2F_{iU}^2F_{ij}^2$ (which shall be fulfilled by our ansatz),
\bea \label{factorized}
{\rm det}(g+2\pi\alpha'{\cal F}) &=& U^8\left(\frac{R}{U}\right)^{3/2}
\left[1+f_T(2\pi\alpha')^2F_{iU}^2+\left(\frac{U}{R}\right)^3 f_T (\partial_U X_4)^2 +(2\pi\alpha')^2\hat{F}_{0U}^2\right] \non[2ex]
&&\times \left[1+\left(\frac{R}{U}\right)^3\frac{(2\pi\alpha')^2F_{ij}^2}{2}\right] \, .
\eea 
The prescription then requires us to expand the square root over this determinant to all orders in $\alpha'$, apply the symmetrized trace for each term separately, and then resum the resulting infinite series. 
Within our ansatz, this resummation
can be done analytically and leads to a relatively simple analytic form for the DBI action, see Sec.\ \ref{sec:str}. Nevertheless, the numerical evaluation turns out to be more difficult compared to 
the simpler prescription that uses the standard (unsymmetrized) trace \cite{Rozali:2007rx,Li:2015uea}. Thus, after showing numerically in Sec.\ \ref{sec:tseytlin} that the results of the two prescriptions do not differ much
we shall resort to the unsymmetrized prescription in Sec.\ \ref{sec:fix}.

\subsubsection{General form of non-abelian field strengths}
\label{sec:general}

The ansatz for the non-abelian part is best discussed in the new holographic coordinate $Z$,
defined as\footnote{Later we shall come back to using the coordinate $U$ in many equations. This is partly to connect to previous literature, and partly because of convenience. Neither 
of the two coordinates turns out to be overly superior when it comes to compactness in notation or convenience in the calculation.}
\be \label{UZ}
U=(U_c^3+U_cZ^2)^{1/3} \, , \qquad \frac{\partial U}{\partial Z} = \frac{2U_c^{1/2}\sqrt{f_c(U)}}{3U^{1/2}} \, , 
\ee
where 
\be
f_c(U) = 1-\frac{U_c^3}{U^3} \, , 
\ee
such that $Z=0$ corresponds to the tip of the connected flavor branes, and $Z=\pm \infty$ to the holographic boundary on the D8- and $\overline{\rm D8}$ branes, see Fig.\ \ref{fig:cylinders}.  
The most general ansatz that is $SO(3)$ symmetric in the spatial directions for an instanton located at $\vec{X}=0$ can be written as \cite{Cherman:2011ve,Rozali:2013fna,Bolognesi:2013jba} 
\begin{subequations} \label{Ageneral}
\bea
A_Z^a(\vec{X},Z) &=& a_Z \frac{X_a}{2X} \, , \\[2ex]
A_i^a(\vec{X},Z) &=& \frac{X\phi_1\delta_{ia}-(1+\phi_2)\epsilon_{ija} X_j}{2X^2} + (Xa_X-\phi_1)\frac{X_iX_a}{2X^3} \, , \label{Aia}
\eea
\end{subequations}
where $X=|\vec{X}|$, and $a_Z$, $a_X$, $\phi_1$, $\phi_2$ are all functions of $X$ and $Z$.
From this ansatz we compute the non-abelian field strengths needed in Eq.\ (\ref{DBIdeconf})  (summation over $a=1,2,3$)
\begin{subequations} 
\bea
F_{iZ}^2 &=& 
\frac{|D_Z\phi|^2}{4X^2} \sigma_a^2 + \frac{F_{XZ}^2 X^2-|D_Z\phi|^2}{4X^4} X_a^2\sigma_a^2
 \, , \\[2ex]
F_{ij}^2 &=& \frac{|D_X\phi|^2}{2X^2}\sigma_a^2 +\frac{(1-|\phi|^2)^2-X^2|D_X\phi|^2}{2X^6}\,X_a^2\sigma_a^2
\, , \\[2ex]
F_{ij}F_{kZ}\epsilon_{ijk} &=& -\frac{{\rm Im}[D_X\phi(D_Z\phi)^*]}{2X^2}\sigma_a^2+\frac{F_{XZ}(1-|\phi|^2)+{\rm Im}[D_X\phi(D_Z\phi)^*]}{2X^4}X_a^2\sigma_a^2\non[2ex]
&&- \frac{i{\rm Re}[D_X\phi(D_Z\phi)^*]}{X^2} \hat{\vec{X}}\cdot \vec{\sigma} \, , \label{Fsq3}
\eea
\end{subequations}
where $\hat{\vec{X}}\equiv \vec{X}/X$, and
\be
\phi= \phi_1+i\phi_2 \, , \quad D_Z = \partial_Z - ia_Z \, , \quad D_X = \partial_X - ia_X\, , \quad F_{XZ} = \partial_X a_Z-\partial_Z a_X \, .
\ee
The field strengths squared are obviously linear combinations of the products 
$\sigma_a\sigma_b$. Except for the non-diagonal structure $\hat{\vec{X}}\cdot \vec{\sigma}$ in $F_{ij}F_{kZ}\epsilon_{ijk}$, there are only diagonal terms, $\sigma_a^2 = 3$, $X_a^2\sigma_a^2 = X^2$.  We have written the Pauli matrices explicitly in the results because this is needed for the discussion of the symmetrized trace, see Sec.\ \ref{sec:str}. 

\subsubsection{Anisotropic instantons}
\label{sec:anisotropic}

We now specify our ansatz for the gauge fields. To this end, it is convenient to work with dimensionless coordinates defined as
\be \label{gammaz}
\vec{x} = \vec{X}M_{\rm KK} \, , \qquad z=\frac{Z}{R(M_{\rm KK}R)^2} \, . 
\ee
In these coordinates, our ansatz for a single instanton placed at $\vec{x}=z=0$ is ($x=|\vec{x}|$)
\begin{subequations} \label{ansatz1}
\bea
a_Z &=& -\frac{1}{R(M_{\rm KK}R)^2}\frac{x}{\gamma}f(x,z) \, , \qquad a_X=M_{\rm KK}\frac{z}{\gamma}f(x,z) \, , \\[2ex]
\phi_1&=&\frac{x z}{\gamma} f(x,z) \, , \qquad \phi_2=x^2f(x,z) -1 \, ,
\eea
\end{subequations}
with 
\be \label{fxz}
f(x,z) = \frac{2}{x^2+(z/\gamma)^2 +(\rho/\gamma)^2} \, .
\ee
This corresponds to the Belavin-Polyakov-Schwarz-Tyupkin (BPST) instanton \cite{1975PhLB...59...85B}, where the $z$ coordinate is rescaled with respect to $x$ by a factor $\gamma$. In appendix 
\ref{app:YM} we derive the single-instanton solution in the deconfined geometry, which does show a nontrivial (and temperature dependent) $\gamma$, see Eq.\ (\ref{gammaT}). Here we will treat $\gamma$ as a 
free parameter, accounting for the "deformation" of the instanton. Such a deformation has also been observed in the 
full solution of a single baryon in the vacuum \cite{Rozali:2013fna,Bolognesi:2013jba}. The instanton width 
in the spatial direction is $\rho/\gamma$, while the width in the holographic direction is $\rho$ (for convenience, we shall often simply refer to $\rho$ as the instanton width). For a given $\rho$, the deformation
parameter thus has the effect of stretching the instanton along the holographic direction $z$ (large $\gamma$) or along the radial spatial direction $x$ (small $\gamma$).
A single instanton becomes elongated along $x$ and wider in both $x$ and $z$ for values of the 't Hooft coupling $\lambda$ away from infinity, which was shown in a full numerical evaluation
of the equations of motion, based on the most general ansatz for the gauge fields (\ref{Ageneral}), see Ref.\ \cite{Rozali:2013fna}. (Already from the SO(4) symmetric case we know that only the finiteness of 
$\lambda$ prevents an instanton in the vacuum from being pointlike \cite{Hata:2007mb}.) Translated to our parametrization, we thus expect a smaller $\gamma$ and a larger $\rho$ for finite $\lambda$ than for $\lambda=\infty$.
Our approximation, based on the ansatz (\ref{ansatz1}), and extended to a many-instanton system below, is too simplistic to allow for a dependence on $\lambda$ 
apart from a trivial rescaling (this is in contrast to the ``homogeneous ansatz'' \cite{Rozali:2007rx,Li:2015uea}, which is not based on any instanton solution). Therefore, besides
computing $\rho$ and $\gamma$ dynamically in Sec.\ \ref{sec:tseytlin}, we shall impose certain constraints on $\rho$ and $\gamma$ in Sec.\ \ref{sec:fix}, with the idea of capturing some
of the $\lambda<\infty$ physics, which seems to be crucial to obtain more realistic results, already for baryons in the vacuum \cite{Cherman:2011ve,Rozali:2013fna,Bolognesi:2013jba}.

With the ansatz (\ref{ansatz1}), the field strengths become particularly simple. The non-diagonal term and all terms proportional to $X_a\sigma_a$ vanish,
\bea \label{zero}
    {\rm Re}[D_X\phi(D_Z\phi)^*] &=& F_{XZ}^2 X^2-|D_Z\phi|^2=(1-|\phi|^2)^2-X^2|D_X\phi|^2 \non[2ex]
    &=&F_{XZ}(1-|\phi|^2)+{\rm Im}[D_X\phi(D_Z\phi)^*]=0 \, ,
\eea
and the remaining terms become proportional to the same function of $x$ and $z$,
\bea \label{relations}
M_{\rm KK}R^3\gamma {\rm Im}[D_X\phi(D_Z\phi)^*] &=& \frac{|D_X\phi|^2}{M_{\rm KK}^2} = M_{\rm KK}^4R^6\gamma^2|D_Z\phi|^2 = \frac{x^2\rho^4f^4(x,z)}{\gamma^4} \, .
\eea
The field strengths now fulfill the relation $(F_{ij}F_{kZ}\epsilon_{ijk})^2=2F_{iZ}^2F_{ij}^2$, and thus we may use Eq.\ (\ref{factorized}). This leads to the DBI action 
\be \label{SDBI1}
S_{\rm DBI} = {\cal N} \int_0^{1/T}d\tau\int d^3X\int_{u_c}^\infty du\, u^{5/2} {\rm str}\sqrt{\left(1+u^3f_T x_4'^2-\hat{a}_0'^2+\frac{g_1\sigma_a\sigma_a}{3}\right)\left(1+\frac{g_2\sigma_a\sigma_a}{3}\right)} \, , 
\ee
where we have replaced $\hat{A}_0 \to i \hat{A}_0$ since we work in Euclidean space, have introduced the dimensionless quantities 
\be \label{smalla0}
\hat{a}_0 = \frac{2\pi\alpha'}{R(M_{\rm KK}R)^2}\hat{A}_0 \, , \qquad  x_4 = M_{\rm KK} X_4 \, , \qquad u=\frac{U}{R(M_{\rm KK}R)^2}\, , 
\ee
and have denoted  the derivative with respect to $u$ by a prime. Also, we have
abbreviated ${\cal N}\equiv 2T_8V_4 R^5(M_{\rm KK}R)^7/g_s$ and 
\begin{subequations} \label{g1g2}
\bea
g_1 &=&
\frac{16\pi^2}{\lambda^2} \frac{f_T}{\gamma^2}\left(\frac{\partial z}{\partial u}\right)^2\frac{12 (\rho/\gamma)^4}{[x^2+(z/\gamma)^2 +(\rho/\gamma)^2]^4}\, , \\[2ex]
g_2 &=&
\frac{16\pi^2}{\lambda^2} \frac{12 (\rho/\gamma)^4}{u^3[x^2+(z/\gamma)^2 +(\rho/\gamma)^2]^4} \, ,
\eea
\end{subequations}
and used the relation $\lambda \ell_s^2 = 2R^3M_{\rm KK}$.

\subsubsection{Spatial average and instanton layers}
\label{sec:spatial}

Next, we go from a single instanton to a many-instanton system. We do so on the level of the field strengths squared.
We place $i_n$ many instantons at the points $z_n$ in the bulk, $n=0,\ldots,N_z-1$ ($N_z\ge 1$),
and distribute them at the points $\vec{x}_{in}$, $i=1,\ldots,i_n$, in position space. The total number of instantons is $N_I\equiv i_0 + \ldots + i_{N_z-1}$. One can think of an instanton
lattice sitting at each point $z_n$ in the bulk. In this general notation, the lattice structure is allowed to be different at different points in the bulk. However, we shall
drastically simplify this general picture in our calculation such that the lattice structure in position space becomes irrelevant: we shall average the field strengths squared over position space
before solving the equations of motion \cite{Ghoroku:2012am,Li:2015uea}, and as a consequence it does not matter at which points
$\vec{x}_{in}$ the instantons sit. The many-instanton system within our approximation is thus obtained by replacing 
\bea \label{Fsq}
\frac{12 (\rho/\gamma)^4}{[x^2+(z/\gamma)^2 +(\rho/\gamma)^2]^4} &\to&  \frac{1}{V}\sum_{n=0}^{N_z-1}\sum_{i=1}^{i_n}\int d^3 X
\frac{12 (\rho/\gamma)^4}{[(\vec{x}-\vec{x}_{in})^2+(z-z_n)^2/\gamma^2 +\rho^2/\gamma^2]^4}
\non[2ex] &=& \frac{2\pi^2\gamma}{M_{\rm KK}^3}\frac{N_{\vec{x}}N_z}{V} \int d^3 x \,D(x,z) = \frac{2\pi^2\gamma}{M_{\rm KK}^3}\frac{N_{\vec{x}}N_z}{V} D(z) \, , 
\eea
where, in the second step, we have assumed that the same number of instantons sits at every $z_n$ and denoted this number by $N_{\vec{x}}$, such that the total number of instantons is now $N_I=N_zN_{\vec{x}}$, 
and where we have defined the normalized instanton profiles
\begin{subequations} 
\bea
D(x,z) &=& \frac{6}{\pi^2\gamma N_z}\sum_{n=0}^{N_z-1}\frac{(\rho/\gamma)^4}{[x^2+(z-z_n)^2/\gamma^2+(\rho/\gamma)^2]^4} \, ,  \label{Dxz} \\[2ex]
D(z) &=& \int d^3 x \,D(x,z) = \frac{1}{N_z}\sum_{n=0}^{N_z-1} \frac{3\rho^4}{4[(z-z_n)^2+\rho^2]^{5/2}} \, , \label{Dz}
\eea
\end{subequations}
with
\be
\qquad \int_{-\infty}^\infty dz \, D(z) = 1 \, .
\ee
It is important that the deformation parameter $\gamma$ has not dropped out, although we have averaged over position space. 
We can thus later compute the deformation of the instantons even though our simplified equations of motion only involve the holographic coordinate $z$, and not $x$. 

\begin{figure}[tbp]
\centering 
\includegraphics[width=.5\textwidth]{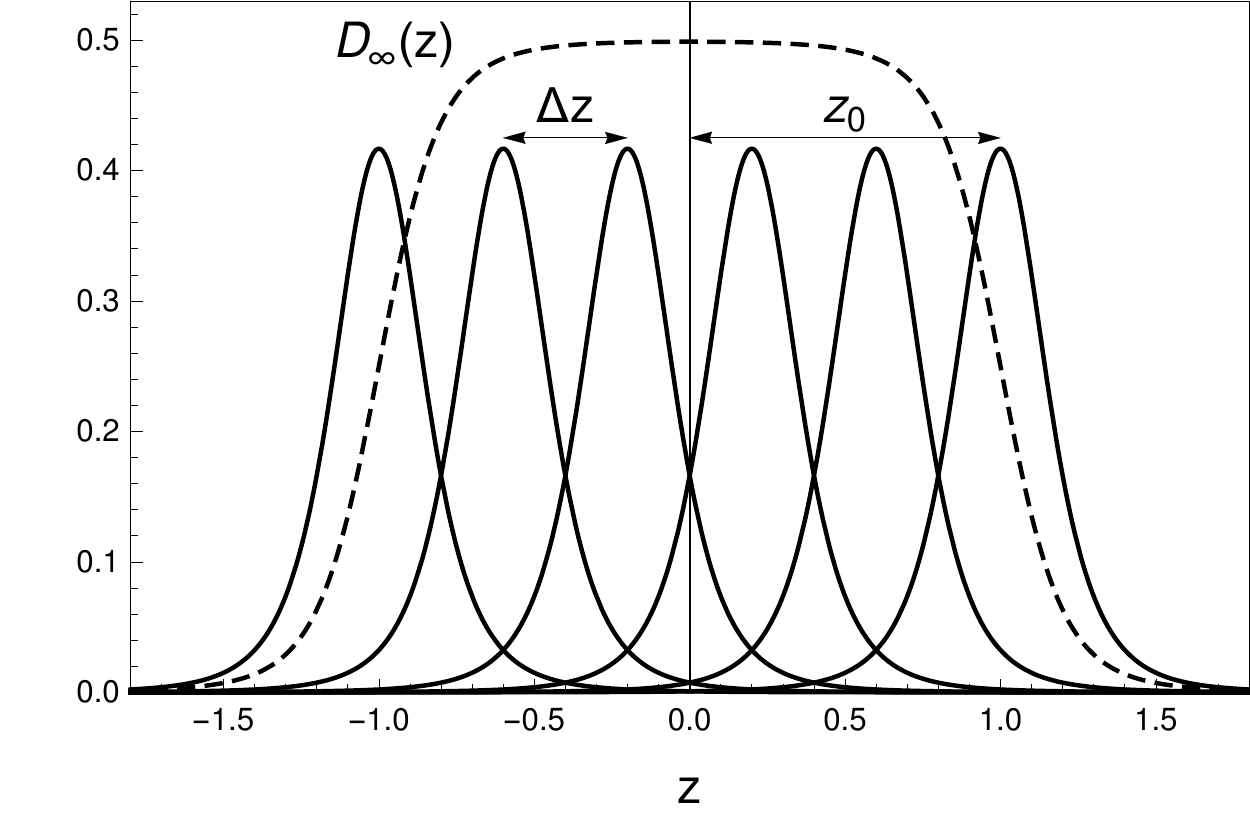}
\caption{\label{fig:many} Instanton distribution along the holographic direction $z$.
 The number of layers $N_z$ and their extension $z_0$ will be determined dynamically. The solid lines are the separate terms in the sum of Eq.\ (\ref{Dz}).
  For $N_z\to\infty$ at fixed $z_0$, the sum over all instanton layers approaches the function $D_\infty(z)$ (\ref{Dzinf}), here shown as a dashed line.
Each instanton layer in the holographic coordinate accommodates $i_n$ instantons in position space, and we assume 
$i_0=\ldots=i_{N_z-1}\equiv N_{\vec{x}}$, the total instanton number thus being $N_I = N_{\vec{x}} N_z$.}
\end{figure}

For the instanton distribution in the bulk we employ the following ansatz.
We assume the layers of $N_{\vec{x}}$ instantons to be separated equidistantly by a distance $\Delta z$ from each other, and to be centered at the points 
\be \label{zn}
z_n= \left(1-\frac{2n}{N_z-1}\right) z_0 \, ,
\ee
such that they are spread over a symmetric interval of length $2z_0$ around the tip of the connected flavor branes $z=0$, and $z_0=(N_z-1)\Delta z/2$. This is illustrated in Fig.\ \ref{fig:many}. This ansatz, where the instanton layers all have the same shape and are separated by the same distance, allows for a continuous transition 
between $N_z=1$ and $N_z=2$, but all other transitions -- if they occur -- will necessarily be discontinuous.

Inserting Eq.\  (\ref{Fsq}) into Eqs.\ (\ref{g1g2}) yields 
\begin{subequations} \label{g1g2sim}
\bea
g_1 &\simeq& \frac{f_T n_I}{3\gamma}\frac{\partial z}{\partial u}q(u) \, , \\[2ex]
g_2 &\simeq& \frac{\gamma n_I}{3u^3} \frac{\partial u}{\partial z} q(u)  \, ,
\eea
\end{subequations}
where we have introduced the dimensionless instanton density (per flavor, hence the division by $N_f=2$)
\be \label{nIdimless}
n_I =  \frac{96\pi^4}{\lambda^2 M_{\rm KK}^3N_f} \frac{N_I}{V}  \, , 
\ee
and defined 
\be
q(u) = 2\frac{\partial z}{\partial u} D(z) \, , \qquad \int_{u_c}^\infty du\, q(u) = 1 \, .
\ee
[Recall that $u=(u_c^3+u_cz^2)^{1/3}$, see Eq.\ (\ref{gammaz}).]
For $N_z=1$ we recover the function $q(u)$ from Ref.\ \cite{Li:2015uea}, where the instanton repulsion was neglected, 
\be
N_z=1:\qquad  q(u) = \frac{9u^{1/2}}{4\sqrt{f_c}}\frac{(\rho^2u_c)^2}{(u^3-u_c^3+\rho^2u_c)^{5/2}} \, .
\ee
We shall treat $z_0$ and $N_z$ as dynamical parameters with respect to which we minimize the free energy. We include the possibility of infinitely many instanton layers, i.e., a smoothly smeared 
instanton distribution. In this limit, letting $N_z\to \infty$ while keeping $z_0$ fixed, we can approximate the sum in Eq.\ (\ref{Dz}) by an integral and we obtain 
\be \label{Dzinf}
D_\infty(z)\equiv \frac{1}{8z_0}\left\{\frac{(z+z_0)[3\rho^2+2(z+z_0)^2]}{[(z+z_0)^2+\rho^2]^{3/2}}-\frac{(z-z_0)[3\rho^2+2(z-z_0)^2]}{[(z-z_0)^2+\rho^2]^{3/2}}\right\} \, ,
\ee
which is also shown in Fig.\ \ref{fig:many}.

\subsubsection{Symmetrized trace}
\label{sec:str}

As explained above, we need to decide on a certain prescription to evaluate the non-abelian DBI action.  
We expand the square root and take the symmetrized trace in each term separately \cite{Tseytlin:1999dj}. 
It is known from string theory that this prescription is accurate up to ${\cal O}(F^4)$ \cite{Sevrin:2001ha}.
For the structure we have in Eq.\ (\ref{SDBI1}), this yields \cite{Kaplunovsky:2010eh}
\bea \label{str}
&&{\rm str}\sqrt{(1+\varphi\sigma_a\sigma_a)(1+\psi\sigma_a\sigma_a)} \non[2ex]
&&= {\rm str}[{\bf 1}]+(\varphi+\psi)\frac{{\rm str}[\sigma_a\sigma_a]}{2}
-(\varphi-\psi)^2\frac{{\rm str}[(\sigma_a\sigma_a)^2]}{8} +(\varphi-\psi)^2(\varphi+\psi)\frac{{\rm str}[(\sigma_a\sigma_a)^3]}{16}+\ldots \non[2ex]
&&=2\left[1+\frac{3}{2}(\varphi+\psi)-\frac{5}{8}(\varphi-\psi)^2+\frac{7}{16}(\varphi-\psi)^2(\varphi+\psi)+\ldots\right]\non[2ex]
&&= 2\frac{(1+2\varphi)(1+2\psi)-\varphi\psi}{\sqrt{(1+\varphi)(1+\psi)}} \, . 
\eea
The equations of motion in this prescription as well as the stationarity equations for the free energy are worked out appendix \ref{app:tseytlin}.
If we instead take the standard trace in this series, we obtain 
\bea \label{naive}
2\sqrt{(1+3\varphi)(1+3\psi)} = 2\left[1+\frac{3}{2}(\varphi+\psi)-\frac{9}{8}(\varphi-\psi)^2+\frac{27}{16}(\varphi-\psi)^2(\varphi+\psi)+\ldots\right] \, . \hspace{0.5cm}
\eea
Since $\varphi, \psi \propto F^2$, this result differs from Eq.\ (\ref{str}) starting from terms of order $F^4$. 
We thus expect different results for large densities. Below we shall present a comparison of the two prescriptions, showing that there is indeed a difference. 
However, this difference turns out to be small and the results are qualitatively the same, see Fig.\ \ref{fig:nIuc}. Therefore, we shall mostly (in the equations in the main text and in all results except for Fig.\ \ref{fig:nIuc}) 
use the unsymmetrized prescription, which leads to significantly simpler equations, resulting in much faster numerics.

\subsection{Chern-Simons action and full Lagrangian}
\label{sec:CS}

Within our ansatz, the CS action is
\bea
S_{\rm CS} &=& \frac{N_c}{8\pi^2}\int_0^{1/T}d\tau \int d^3 X \int_{-\infty}^\infty dZ\,\hat{A}_0\Tr[F_{ij}F_{kZ}]\epsilon_{ijk} \, .
\eea
Having computed the field strengths and having introduced convenient dimensionless quantities, we can easily compute this contribution. We obtain $F_{ij}F_{kZ}\epsilon_{ijk}$ from Eqs.\ (\ref{Fsq3})
(\ref{zero}), and (\ref{relations}), then use Eq.\ (\ref{Fsq}) as well as $\Tr[\sigma_a\sigma_a]=6$ to obtain
\be
S_{\rm CS} = -{\cal N}\frac{V}{T}N_f n_I\int_{u_c}^\infty du\,\hat{a}_0 q(u) \, .
\ee
Putting this together with the DBI action in the unsymmetrized prescription, this yields the action 
\be \label{Sfinal}
S = {\cal N} \frac{V}{T} N_f \int_{u_c}^\infty du\, {\cal L} \, ,
\ee
with the Lagrangian 
\be \label{Lag}
{\cal L} = u^{5/2} \sqrt{(1+u^3f_T x_4'^2-\hat{a}_0'^2+g_1)(1+g_2)} - n_I\hat{a}_0 q(u) \, ,
\ee
with $g_1$ and $g_2$ from Eqs.\ (\ref{g1g2sim}).
This Lagrangian has exactly the same form as the one used in Ref.\ \cite{Li:2015uea}, see Eq.\ (30) in that reference. The extensions of the present approach are hidden in the functions $g_1$, $g_2$:
we reproduce the functions $g_1$, $g_2$ of Eq.\ (31) in Ref.\ \cite{Li:2015uea} by considering only one instanton layer, $N_z=1$, $z_0=0$, and by choosing  the instanton deformation to be $\gamma= 3u_c^{3/2}/2$
(this specific value was chosen by transferring the BPST result of the confined geometry to the deconfined
geometry)\footnote{In Ref.\ \cite{Li:2015uea} the flavor trace in the DBI action was performed {\it only} over the $F^2$ terms, in apparent disagreement with the prescription (\ref{naive}).
  However, this merely leads to a redefinition of $n_I$, which was defined {\rm without} including a factor $1/N_f$ and to the absence of the overall prefactor of the action $N_f$.
  Since $n_I$ is a dynamical quantity, determined by minimizing the free energy, this difference does not matter.}.

\subsection{Minimizing the free energy}
\label{sec:mini}

The equations of motion for $\hat{a}_0$ and $x_4$, obtained from the Lagrangian (\ref{Lag}), are, in integrated form, 
\begin{subequations}\label{eoms}
\bea
\frac{u^{5/2}\hat{a}_0'\sqrt{1+g_2}}{\sqrt{1+g_1+u^3f_Tx_4'^2-\hat{a}_0'^2}}&=& n_I Q \, , \label{eom1}\\[2ex]
\frac{u^{5/2}u^3f_Tx_4'\sqrt{1+g_2}}{\sqrt{1+g_1+u^3f_Tx_4'^2-\hat{a}_0'^2}}&=& k \, , \label{eom2}
\eea
\end{subequations}
where $k$ is an integration constant, and 
\bea \label{Qdef}
Q(u) &\equiv& \int_{u_c}^u dv\,q(v) = \int_{-z(u)}^{z(u)} dy\,D(y) \non[2ex]
&=&\frac{1}{N_z}\sum_{n=0}^{N_z-1}\left\{\frac{[2(z-z_n)^2+3\rho^2](z-z_n)}{4[(z-z_n)^2+\rho^2]^{3/2}}+ \frac{[2(z+z_n)^2+3\rho^2](z+z_n)}{4[(z+z_n)^2+\rho^2]^{3/2}}\right\} \, .
\eea
For $N_z\to \infty$ we have
\be
Q_\infty(z) 
=\frac{1}{4z_0}\left[\frac{\rho^2+2(z+z_0)^2}
{\sqrt{\rho^2+(z+z_0)^2}} -\frac{\rho^2+2(z-z_0)^2}
{\sqrt{\rho^2+(z-z_0)^2}}\right] \, .
\ee
The equations of motion (\ref{eoms}) can easily be solved for $\hat{a}_0'$ and $x_4'$ algebraically. The resulting expressions can then be inserted into the
(dimensionless) free energy density of the baryonic phase,
\bea \label{OmegaTV}
\Omega_{\rm baryon} &\equiv& \int_{u_c}^\infty du\, {\cal L} \non[2ex]
&=& \int_{u_c}^\infty du\,u^{5/2} \sqrt{1+g_1}\sqrt{1+g_2-\frac{k^2}{u^8f_T}+\frac{(n_IQ)^2}{u^5}} +\frac{\ell}{2}k-\mu n_I \, ,
\eea
where the Lagrangian is given in Eq.\ (\ref{Lag}) and where, in the second line, we have employed partial integration in the CS term and used the boundary conditions $\hat{a}_0(\infty)=\mu$, $x_4(\infty)=\ell/2$, with the dimensionless chemical potential $\mu$ and the dimensionless asymptotic separation of the flavor branes $\ell = M_{\rm KK}L$. (The complete, dimensionful free energy density is obtained by multiplying $\Omega_{\rm baryon}$ with ${\cal N}N_f$.)
We note that the asymptotic behavior of $\hat{a}_0'$ and $x_4'$ is given by
\be
x_4'(u) = \frac{k}{u^{11/2}} + \ldots \, , \qquad \hat{a}_0'(u) = \frac{n_I}{u^{5/2}} + \dots \, . 
\ee
This confirms that $n_I$ is the (dimensionless) baryon density, which is also given by the derivative of the free energy with respect to the chemical potential,
\be\label{nI}
n_I = -\frac{\partial \Omega_{\rm baryon}}{\partial \mu} \, . 
\ee
This equation seems like an obvious thermodynamic relation, but there are some subtleties in the Sakai-Sugimoto model if baryon number is (partially) created through a magnetic field 
\cite{Bergman:2008qv,Rebhan:2009vc}. In that case, a modified Chern-Simons term has been used to 
ensure the relation (\ref{nI}) \cite{Bergman:2008qv,Lifschytz:2009sz,Rebhan:2009vc,Preis:2010cq,Preis:2011sp}. 
Here, no such modification is necessary. 

To be precise about the meaning of our dimensionless quantities, we notice that $\mu$ is a dimensionless {\it quark} chemical potential, while $n_I$ is a dimensionless {\it baryon} number density.
The physical quark chemical potential is related to $\mu$ by the factor introduced in the definition of the dimensionless 
abelian gauge field in Eq.\ (\ref{smalla0}),
\bea\label{dimful1}
\mbox{quark chemical potential} &=& \frac{\lambda M_{\rm KK}}{4\pi} \mu \, . 
\eea
Inserting this relation into Eq.\ (\ref{nI}) and using that the dimensionful free energy density is ${\cal N}N_f\Omega_{\rm baryon}$, we read off the physical quark number density, i.e., 
\bea\label{dimful2}
\mbox{quark number density} &=& N_cN_f\frac{\lambda^2M_{\rm KK}^3}{96\pi^4} n_I \, .
\eea
With Eq.\ (\ref{nIdimless}) we conclude that the dimensionful baryon number density (= quark number density divided by $N_c$) is exactly the instanton density $N_I/V$.

The free energy (\ref{OmegaTV}) is a function of the parameters $k$, $n_I$, $u_c$, $\rho$, $\gamma$, $z_0$, $N_z$. They are independent of each other except for the obvious condition that the separation of 
instanton layers vanishes, $z_0=0$, if and only if there is exactly one instanton layer, $N_z=1$.  
We shall discuss the following two approaches and present their results in
Secs.\ \ref{sec:tseytlin} and \ref{sec:fix}, respectively. 

\begin{enumerate}

\item[(i)] Minimize $\Omega_{\rm baryon}$ with respect to all seven parameters. 
  
\item[(ii)] Impose the following constraints on the parameters that determine the shape of a single instanton, i.e., the instanton width $\rho$ and instanton deformation $\gamma$,
\be\label{rhogamma}
\rho = \rho_0 u_c \, , \qquad \gamma = \frac{3}{2}\gamma_0u_c^{3/2} \, ,
\ee
and fix $\rho_0$, $\gamma_0$. Then minimize $\Omega_{\rm baryon}$ with respect to the remaining five parameters $k$, $n_I$, $u_c$, $z_0$, $N_z$. 
\end{enumerate}

Approach (i) requires no further motivation, it yields the ground state that the system chooses to be in within the given approximation.
The idea behind approach (ii) is as follows. We do not know how our many-instanton ansatz is related to the full solution of the problem. Therefore, we have to take the result of the straightforward
minimization of scenario (i) with some care: the minimum in our restricted parameter space might be very different from the minimum in the full functional space. However,
as mentioned below Eq.\ (\ref{fxz}),
we do know some features of the full solution of single instantons in the vacuum, in particular we know that the width and the deformation change as a function of $\lambda$ away from the $\lambda=\infty$ limit.
In order to consider a many-instanton system, we have given up some complexity, in particular we do not expect our ansatz to reproduce these important $\lambda <\infty$  features of single instantons. 
Therefore, we choose to impose external constraints on width and deformation and scan through the resulting parameter space. 
We might simply have "rigidly" fixed $\rho$ and $\gamma$. However, we do expect these quantities to change with density.
Therefore, we have chosen a particular scaling with (the density-dependent) $u_c$. This "natural" scaling is chosen such that $u_c$ drops out of but one minimization equations, as we shall see below.
(The factor 3/2 in the scaling relation for $\gamma$ is chosen such that $\gamma_0=1$ corresponds to the calculation done in Ref.\ \cite{Li:2015uea}.) The use of the constraints on $\rho$ and $\gamma$ is 
justified a posteriori by the observation that only in approach (ii) we do find a layered structure of the instantons in the bulk, whose existence is suggested from other, complementary, approximations in the literature.  

In both approaches (i) and (ii), the parameters are determined by setting the various derivatives of the free energy to zero. 
The parameter $N_z$ is special because it is an integer and thus we cannot simply take the derivative of the free energy with respect to $N_z$. Instead, we will solve the below equations 
for various values of $N_z$, including $N_z=\infty$. It turns out that there is a clear tendency in the behavior of the free energy as a function of $N_z$, and thus this procedure is sufficient to determine the preferred $N_z$. 
Since we have written the free energy in the same form as in Ref.\ \cite{Li:2015uea}, we can skip the details of the derivation of the stationarity equations and directly quote the
results. Anyway, only the derivative with respect to $u_c$ is not completely straightforward, see Sec.\ III of Ref.\ \cite{Li:2015uea} or appendix \ref{app:tseytlin} of the present paper, where we go into some details in the context of the symmetrized trace prescription. 
The resulting equations (in the order:  minimization with respect to 
$k$, $n_I$, $\rho$, $\gamma$, $z_0$, $u_c$) are
\begin{subequations} \label{mini}\allowdisplaybreaks
\bea
\frac{\ell}{2} &=& \int_{u_c}^\infty du\,x_4' \, , \label{ell} \\[2ex]
\mu &=&  \int_{u_c}^\infty du\, \left[\hat{a}_0'Q+\frac{u^{5/2}}{2}\left(\frac{\partial g_1}{\partial n_I}\zeta^{-1}
+\frac{\partial g_2}{\partial n_I}\zeta \right) \right] \, , \label{mini1}\\[2ex]
0 &=&  \int_{u_c}^\infty du\, \left[\frac{u^{5/2}}{2}\left(\frac{\partial g_1}{\partial \rho}\zeta^{-1}
+\frac{\partial g_2}{\partial \rho}\zeta \right) + n_I \hat{a}_0'\frac{\partial Q}{\partial \rho}\right] \, ,\label{mini2} \\[2ex]
0 &=&  \int_{u_c}^\infty du\, u^{5/2}(-g_1\zeta^{-1}+g_2\zeta)\, ,\label{mini3} \\[2ex]
0 &=&  \int_{u_c}^\infty du\, \left[\frac{u^{5/2}}{2}\left(\frac{\partial g_1}{\partial z_0}\zeta^{-1}
+\frac{\partial g_2}{\partial z_0}\zeta \right) + n_I \hat{a}_0'\frac{\partial Q}{\partial z_0}\right] \, ,\label{mini4} \\[2ex]
0 &=& \int_{u_c}^\infty du 
\left[\frac{u^{5/2}}{2}(g_1\zeta^{-1}p_-+g_2\zeta p_+)+n_I \hat{a}_0'\frac{\partial Q}{\partial u_c}-\frac{\alpha k(u-u_c)^{-3/2}}{6u_c^2 \gamma _0c_1}+\frac{3u_c^2}{u^{1/2}f_c}\frac{g_1}{2\zeta} \right] \, , \hspace{0.9cm}\label{mini5}
\eea
\end{subequations}
where we have used the abbreviation
\be \label{defzeta0}
\zeta \equiv \frac{\sqrt{1+g_1}}{\sqrt{1+g_2-\frac{k^2}{u^8f_T}+\frac{(n_IQ)^2}{u^5}}} \, , 
\ee
and where, in Eq.\ (\ref{mini5}), we have defined
\be 
p_\pm \equiv \frac{1}{q\sqrt{f_c}}\frac{\partial(q\sqrt{f_c})}{\partial u_c}\pm\frac{2}{u_c} \, , \quad  c_1 \equiv \frac{\alpha^{1/2}k}{u_c\sqrt{3\gamma_0}\sqrt{(1+\gamma_0\alpha)f_Tu_c^8-k^2}} \, ,
\ee
with $c_1$ giving the behavior of $x_4'$ close to $u_c$, $x_4' = c_1(u-u_c)^{-1/2}+\ldots$, and
\be \label{alpha}
\alpha \equiv  \frac{3n_I}{4u_c^{3/2}}\frac{1}{N_z}\sum_{n=0}^{N_z-1}\frac{\rho^4}{(\rho^2+z_n^2)^{5/2}} \;\underset{N_z\to\infty}{=} \; \frac{n_I(3\rho^2+2z_0^2)}{4u_c^{3/2}(\rho^2+z_0^2)^{3/2}} \, . 
\ee
Notice that the minimization with respect to $k$ (\ref{ell}) is nothing but the condition that the asymptotic separation of the flavor branes be $\ell$. 

We thus have to solve 6 coupled equations for $k$, $n_I$, $u_c$, $\rho$, $\gamma$, $z_0$ in approach (i) and 4 coupled equations for $k$, $n_I$, $u_c$, $z_0$ in approach (ii) (and do so for various values of $N_z$). However, in both cases, two equations decouple.
First, we observe that the only explicit appearance of $\mu$ is in Eq.\ (\ref{mini1}). Therefore, 
rather than fixing $\mu$ we can fix $n_I$ and determine the corresponding $\mu$ with the help of Eq.\ (\ref{mini1}) after we have solved the other equations. (This is also advantageous because 
$\mu$ is always a single-valued function of $n_I$, while $n_I$ can become a multi-valued function of $\mu$.) Second, we can rescale all quantities with appropriate 
powers of $u_c$ and introduce the new integration variable $u/u_c$. One can show that this eliminates $u_c$ from all equations except for Eq.\ (\ref{ell}). Hence,  Eq.\ (\ref{ell})
also decouples, we can solve the remaining equations for the rescaled quantities, then compute $u_c$ from Eq.\ (\ref{ell}) and then undo the rescaling with the help of the resulting $u_c$. 
A similar rescaling of all quantities with the externally given parameter $\ell$ eliminates $\ell$ from all equations. As a consequence, all results scale with $\ell$ in a simple way;
different $\ell$'s do not lead to qualitatively different results. 

Since the rescaling with $u_c$, in particular together with our two approaches (i) and (ii), may be somewhat confusing, let us explain this in more detail.
In deriving Eq.\ (\ref{mini5}) we have {\it first} applied Eq.\ (\ref{rhogamma}) and {\it then} taken the derivative with respect to $u_c$. For scenario (i) this is not very crucial because
minimizing with respect to $u_c$, $\rho$, $\gamma$ is equivalent to minimizing with 
respect to $u_c$, $\rho_0$, $\gamma_0$. [To see this, consider the function $\Omega_{\rm baryon}=\Omega_{\rm baryon}(u_c,\rho,\gamma)$ and take the derivatives with respect to
  $u_c$, $\rho$, $\gamma$ on the one hand and, via the chain rule, with 
respect to $u_c$, $\rho_0$, $\gamma_0$ on the other hand. The apparent additional terms created in the latter procedure are all zero because the derivatives with 
respect to $\rho_0$ and $\gamma_0$ are required to vanish.]
In scenario (ii) it {\it is}  crucial to correctly capture the dependence on $u_c$ within $\rho$ and 
$\gamma$, because we do not minimize with respect to $\rho_0$ and $\gamma_0$. In both scenarios, the eventual rescaling with $u_c$ (where $\rho_0$ and $\gamma_0$ by construction do not scale 
anymore with $u_c$) is then merely a convenient trick to simplify the numerical evaluation. As a check, we have also evaluated the equations without this eventual rescaling and found the same result. 

Once the minimum of the free energy is found within our ansatz for baryonic matter, we need to compare the value of $\Omega_{\rm baryon}$ at that minimum with the free energies of the mesonic phase (= chirally broken phase without 
nuclear matter) and the quark matter phase (= chirally restored phase), see Fig.\ \ref{fig:cylinders}. 
We shall restrict ourselves to zero temperature, although the equations derived above for the baryonic phase provide the full temperature dependence.
At zero temperature, the free energy of all three phases at the stationary point can be written in the very compact form 
\begin{equation}\label{eq:compfreeenergy}
\Omega=\frac{2}{7}\Lambda^{7/2}-\frac{2}{7}\mu n_I-\frac{1}{14}k\ell \, , 
\end{equation}
where $\Lambda$ is an ultraviolet cutoff, replacing the boundary $u=\infty$, and where $n_I$ and $k$ remain to be determined numerically in the baryonic phase and 
have simple analytic forms in the mesonic and quark matter phases, see below. All free energies show the same constant divergence which becomes irrelevant 
when we compare them which each other. In the baryonic phase, the form\ (\ref{eq:compfreeenergy}) is derived as follows. We start from Eq.\ (\ref{OmegaTV}), rescale 
 $n_I\to u_c^{5/2}n_I$, $k\to u_c^4 k$, $\rho\to u_c\rho$, $\gamma\to u_c^{3/2}\gamma$, and introduce the new integration variable $u/u_c$. We do not 
 rescale the externally given quantities $\mu$ and $\ell$. Neither would we rescale $T$, but we have already set $T=0$ and thus $f_T=1$.
Then, we extremize  the resulting expression with respect to $u_c$, taking into account the $u_c$ dependence in the upper boundary of the integral, which now 
is $\Lambda/u_c$. The condition that the derivative of $\Omega_{\rm baryon}$ with respect to $u_c$ vanishes, gives exactly Eq.\ (\ref{eq:compfreeenergy}). As an aside, this procedure also gives an 
alternative form of the minimization with respect to $u_c$ (\ref{mini5}). For the 
mesonic and quark matter phases we use the well-known results (for instance from appendix B of Ref.\ \cite{Li:2015uea}) and note that at $T=0$ they have the form 
(\ref{eq:compfreeenergy}) with 
\begin{subequations}
\bea
\mbox{quark matter:}&& \qquad n_I=\mu^{5/2}\left[\frac{\sqrt{\pi}}{\Gamma\left(\frac{3}{10}\right)\Gamma\left(\frac{6}{5}\right)}\right]^{5/2}\, , \qquad k=0 \, ,  \\[2ex]
\mbox{mesonic phase:}&& \qquad   n_I=0 \, , \qquad k=u_c^4 \, , \quad 
u_c=\left[\frac{4\sqrt{\pi}\Gamma\left(\frac{9}{16}\right)}{\ell\Gamma\left(\frac{1}{16}\right)}\right]^2\,  .  
\eea
\end{subequations}
In particular, the free energy of the quark matter phase does not depend on $\ell$, while the free energy of the mesonic phase does not depend on $\mu$.

\section{Results}
\label{sec:results}

\subsection{Fully dynamical instanton width and deformation}
\label{sec:tseytlin}

\begin{figure}[tbp]
\centering
\includegraphics[width=.7\textwidth]{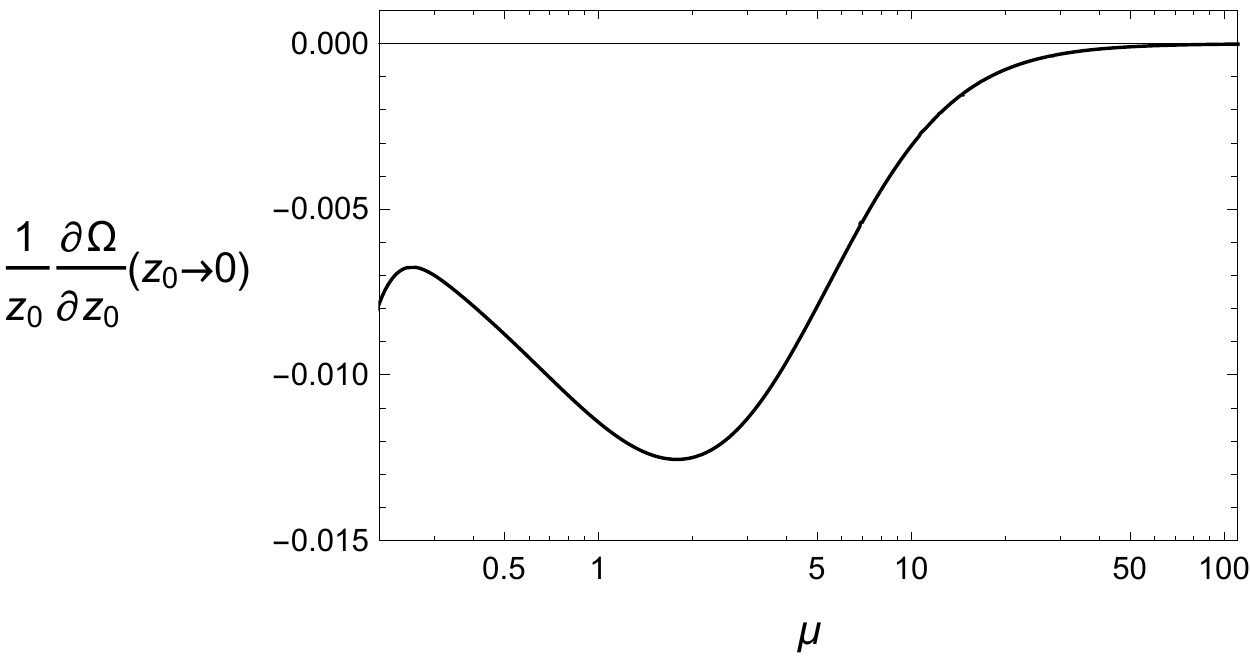}
\caption{\label{fig:absence} Absence of multiple instanton layers in approach (i): 
the plotted function is the right-hand side of Eq.\ (\ref{mini4}), divided by $z_0$, evaluated at $z_0\to 0$, and with $n_I$, $k$, $\rho$, and $\gamma$ 
from the $z_0=0$ solution. [In that limit, $N_z$ still appears in the prefactor, see Eqs.\ (\ref{expand}); without loss of generality, we have set $N_z=\infty$ for this plot.] A zero of this function would give a critical chemical potential at which a solution with more than one layer, $z_0>0$, starts to exist.
The numerical result shows the absence of such a critical chemical potential (and suggests that a layered structure is approached asymptotically at $\mu=\infty$). 
The plot does not exclude the possibility of a discontinuous  transition to $z_0>0$, but we have not found such a transition.}
\end{figure}

In this section, we evaluate and discuss approach (i), i.e., we minimize the free energy with respect to all free parameters, including the instanton width $\rho$ and the deformation $\gamma$. 
For the minimization with respect to $z_0$ (\ref{mini4}) we  observe
that for small $z_0$
\begin{subequations} \label{expand}
\bea
\frac{\partial q}{\partial z_0} &=& -\frac{5}{2}\frac{N_z+1}{N_z-1}\frac{\partial z}{\partial u}\frac{\rho^2-6z^2}{(\rho^2+z^2)^{9/2}} z_0 + {\cal O}(z_0^2) \, , \\[2ex]
\frac{\partial Q}{\partial z_0} &=& -\frac{5}{2}\frac{N_z+1}{N_z-1}\frac{z\rho^4}{(\rho^2+z^2)^{7/2}}z_0 + {\cal O}(z_0^2) \, .
\eea
\end{subequations}
This implies that $z_0=0$ always solves Eq.\ (\ref{mini4}), and thus one solution of our system is always found by solving the remaining equations with $z_0=0$. 
In those equations, then, the number of instanton layers in the bulk $N_z$ only appears implicitly in $n_I\propto N_{\vec{x}}N_z/V$, which is determined dynamically anyway. We thus do not have to compute the free energies for various values of $N_z$ separately to find the ground state.  
This has to be done only for the solutions with $z_0>0$. To search for such solutions let us first assume that there is a continuous transition from $z_0=0$ to $z_0>0$ as a function of $\mu$. 
The critical chemical potential for this transition can be found by dividing Eq.\ (\ref{mini4}) by $z_0$ to exclude the trivial solution and inserting $n_I$, $\rho$, $\gamma$ and $k$ from the $z_0=0$ solution into 
the $z_0\to 0$ limit of the resulting equation.
[This is just like computing the critical temperature of a second-order phase transition with (\ref{mini4}) playing the role of a gap equation.] The expansions (\ref{expand}) show that 
the critical chemical potential (if it exists) does not depend on $N_z$ because $N_z$ only enters in the overall prefactor of the equation (the special case $N_z=1$, in which that prefactor 
diverges, brings us back to the trivial solution $z_0=0$). 
In other words, the points at which the solutions for $N_z\ge 2$ instanton layers start to exist all fall together to a single point. It turns out that in approach (i) this point does not exist. This can be shown numerically by computing the right-hand side of Eq.\ (\ref{mini4}), with $n_I$, $\rho$, $\gamma$ and $k$ from the $z_0=0$ solution inserted. We have plotted the result as a function of $\mu$ in Fig.\ \ref{fig:absence}. A zero of the plotted curve would correspond to a critical chemical potential for the onset of a second instanton layer. We see that there is no zero. Interestingly, 
multiple layers seem to be "postponed" to infinitely large densities because the plotted function approaches zero asymptotically for $\mu\to \infty$.

This argument does not exclude that there is a {\it discontinuous} transition to a phase with multiple instanton layers. In a numerical search 
we have not found any solution $z_0>0$, but a rigorous proof for that absence is difficult.  
We discuss the only solution we have found, $N_z=1$, $z_0=0$, in the following, and come back to 
solutions that show instanton repulsion in Sec.\ \ref{sec:fix}, where we work with approach (ii), in which case we {\it do} find solutions with $z_0>0$, both in a continuous and a discontinuous transition, depending on the values 
of $\rho_0$ and $\gamma_0$.

\begin{figure}[tbp]
\centering 
\hbox{\includegraphics[width=.49\textwidth]{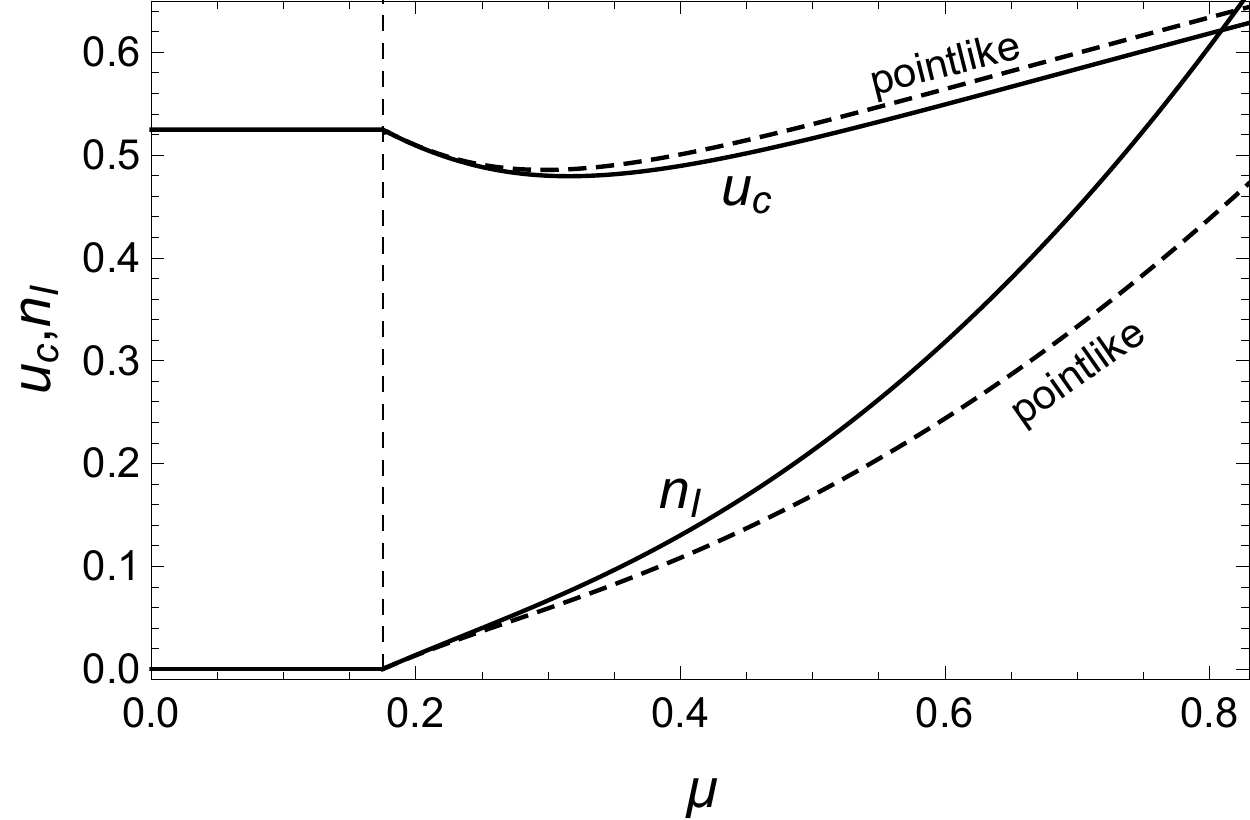}\includegraphics[width=.49\textwidth]{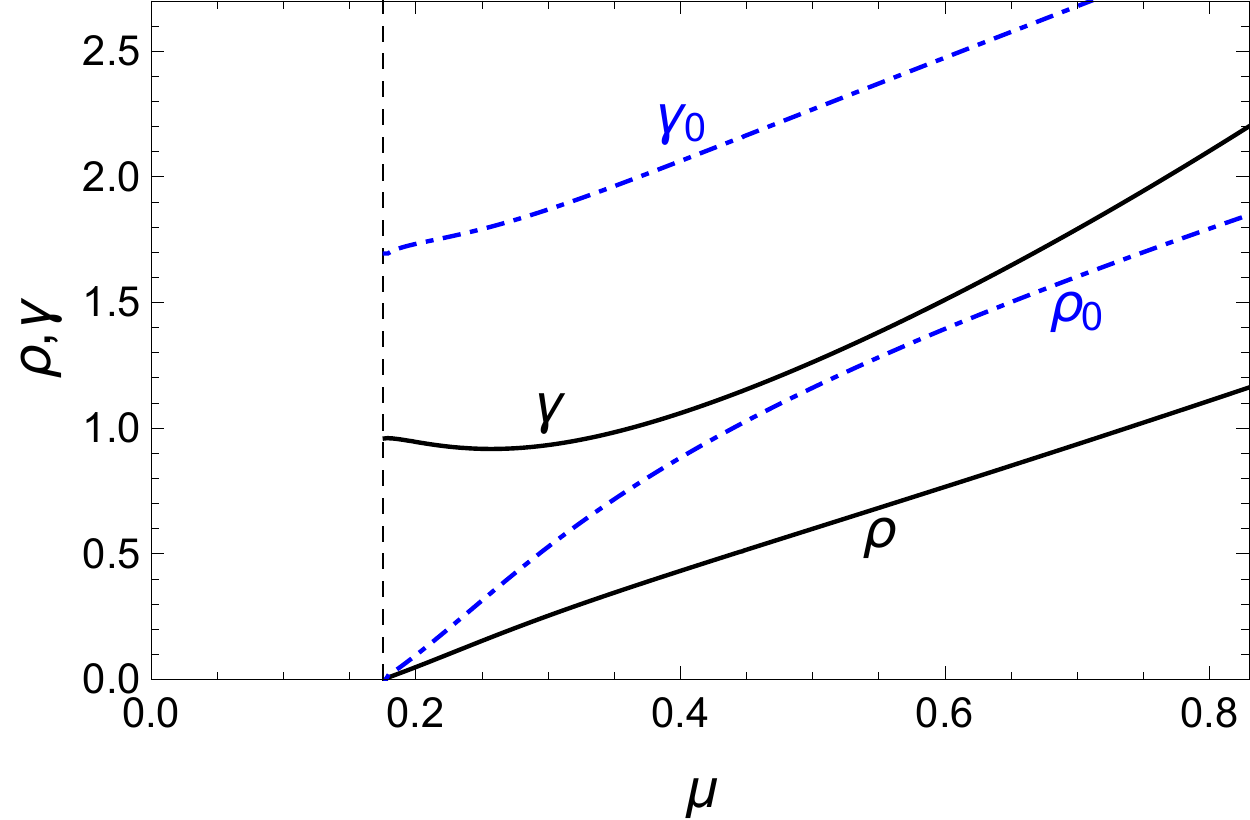}}

\hbox{\includegraphics[width=.49\textwidth]{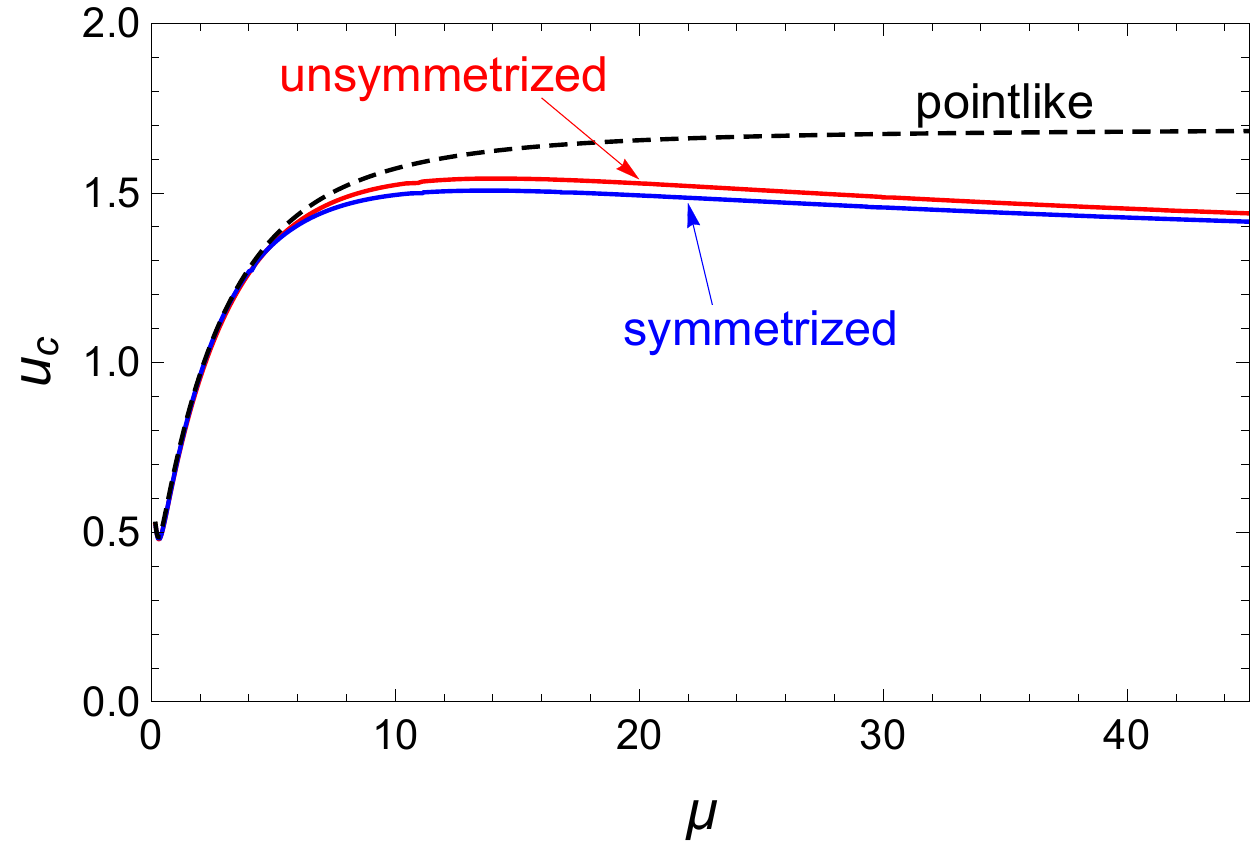}\includegraphics[width=.49\textwidth]{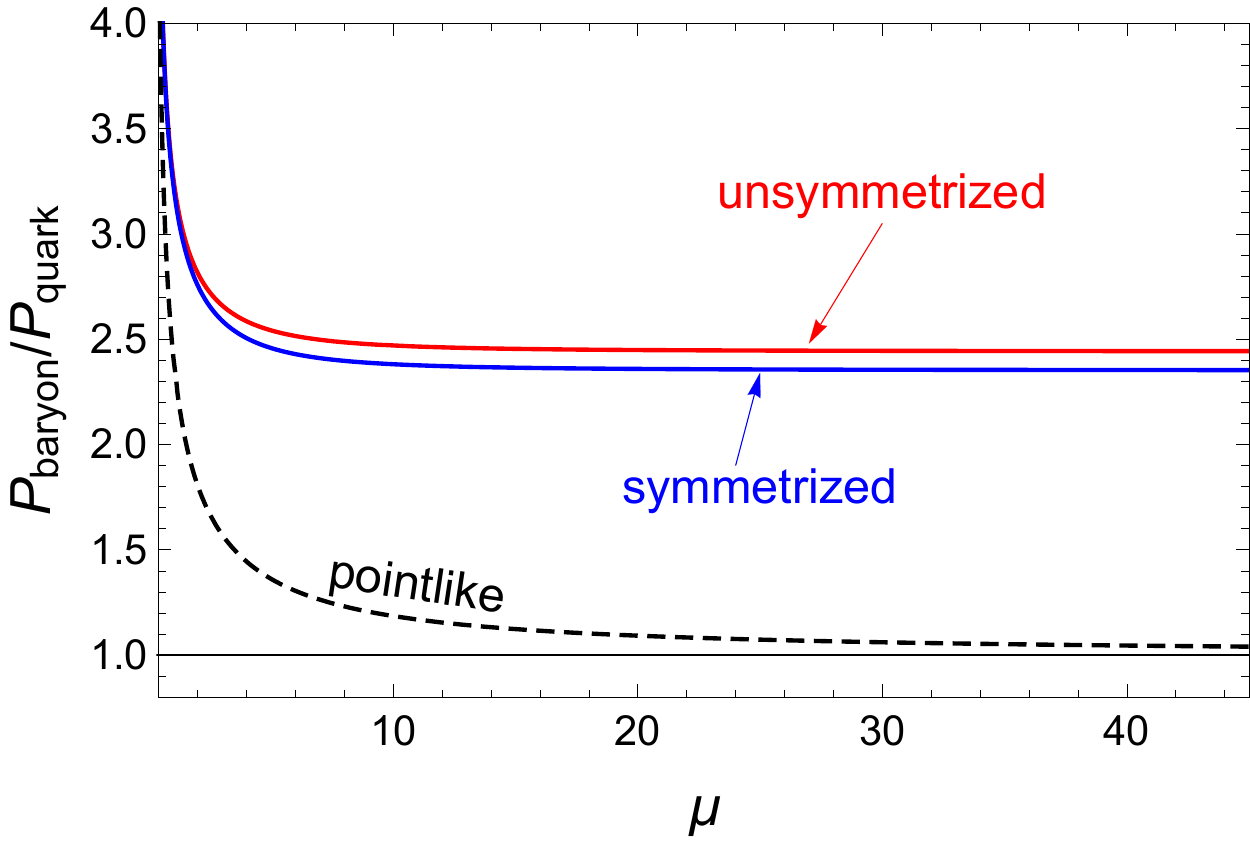}}
\caption{\label{fig:nIuc} Results after minimizing with respect to instanton deformation and instanton width [approach (i)]. Upper left panel: location of the tip of the connected flavor branes $u_c$ and baryon density $n_I$ (solid lines),
compared to the corresponding quantities in the pointlike approximation \cite{Bergman:2007wp} (dashed lines). The vertical dashed line marks the baryon onset at $\mu \simeq 0.175$. Upper right panel: 
instanton deformation $\gamma$ and instanton width $\rho$ (solid lines). We have also plotted $\gamma_0$ and $\rho_0$ (dashed-dotted lines), related to $\gamma$ and $\rho$ via Eq.\ (\ref{rhogamma}),  
which are relevant for a comparison to the results of Sec.\ \ref{sec:fix}, where we work with fixed $\gamma_0$ and $\rho_0$ [approach (ii)]. Lower panels: comparison between the symmetrized trace prescription 
(blue solid lines), the unsymmetrized one (red solid lines) and the pointlike approximation (dashed lines). The lower right panel shows that there is no chiral restoration, 
$P_{\rm baryon}/P_{\rm quark}>1$ for all $\mu$. Note the much larger $\mu$ scale in the lower panels compared to the upper ones. Here and in all other figures, we have set the temperature to zero.}
\end{figure}

The results for $N_z=1$, $z_0=0$ are shown in Fig.\ \ref{fig:nIuc}, which leads to the following observations\footnote{All quantities in this and all following figures are 
 rescaled with appropriate powers of 
$\ell$, i.e., $\mu$ stands for $\ell^2\mu$, $n_I$ for $\ell^5n_I$ etc.  
If we wish to assign physical units to the plot, we have to choose values for the three parameters of the model, say $M_{\rm KK}=950\,{\rm MeV}$, $\lambda =16$, and $L=0.3\,\pi/M_{\rm KK}$. Then, 
using the relations (\ref{dimful1}) and (\ref{dimful2}) for the dimensionful chemical potential and density, the maximum 
baryon chemical potential in the two upper panels is $3.3\, {\rm GeV}$ (about 3.6 times the chemical potential of the real-world baryon onset), while the baryon density at that point is $5.3\,{\rm fm}^{-3}$
(about 35 times real-world nuclear saturation density). Here we are only interested in the qualitative properties of our approximations -- which, in Fig.\ \ref{fig:nIuc}, are not in agreement with real-world baryonic matter -- 
and thus these numbers do not mean much.}.

\begin{itemize}
\item The transition from the mesonic to the baryonic phase is second order, as can be seen from the continuity of the baryon density in the upper left panel. This implies the absence of a binding energy, in contradiction to real-world nuclear matter. This result is qualitatively the same as for the pointlike approximation of baryons (shown as dashed lines). At the baryon onset, our solution 
approaches that of the pointlike approximation. Qualitatively, this is exactly the same observation that had already been made in Ref.\ \cite{Li:2015uea}, where the instanton deformation was not determined dynamically and
where only one instanton layer was taken into account. We thus conclude that allowing for a dynamical instanton deformation and multiple instanton layers
is not sufficient to turn the unphysical second order baryon onset into a physical first order transition.

\item The instanton deformation $\gamma$ decreases just after the onset and then increases for large baryon densities, for very large $\mu$ we find $\gamma\propto \mu^2$. Hence at large densities the instanton becomes elongated along the holographic direction. More specifically, the instanton width in the holographic direction 
$\rho$ increases monotonically with density, with asymptotic behavior $\rho\propto \mu^{3/2}$, while the instanton width in the spatial direction $\rho/\gamma$ behaves non-monotonically, increasing for small $\mu$ 
and decreasing like $\mu^{-1/2}$ for very large $\mu$. At the baryon onset, 
the width of the instanton is zero in all directions. Since the density just above the onset is infinitesimally small, our approximation thus predicts a pointlike baryon in the vacuum. We know that holographic baryons do acquire 
a width if corrections of finite $\lambda$ are taken into account. This indicates that our present approximation is too simplistic to yield realistic isolated baryons. 

\item There is no chiral restoration at large chemical potentials. This can be seen from the lower right panel, where the ratio of the baryonic pressure over the pressure of the chirally restored phase 
(quark matter) is shown ($P=-\Omega$). Chiral restoration would occur if that ratio were to decrease below 1. While in the pointlike approximation this ratio approaches 1 for $\mu\to\infty$ (which can be shown analytically
\cite{Preis:2011sp}), the ratio appears to saturate at a much larger value, $P_{\rm baryon}/P_{\rm quark} \sim 2.4$, for our extended, and deformed, instantons. Again, it is instructive to compare this result to that of Ref.\ \cite{Li:2015uea}, 
where the deformation was fixed. In that case, chiral restoration did occur. Here, we allow the system to settle at a lower free energy by adjusting its instanton deformation. As a consequence, the transition to quark matter has disappeared.  

\item In the lower panels we compare the results for the two different prescriptions for the non-abelian DBI action. We see that they do differ for large chemical potentials, the free energy from the 
symmetrized prescription is somewhat larger (smaller ratio $P_{\rm baryon}/P_{\rm quark}$), however the difference is small and not relevant for our main conclusions. Had we plotted both results in the upper panels, 
the curves would have been indistinguishable by naked eye. Since the unsymmetrized prescription is much simpler, we shall in the following section only work with it and discard the symmetrized prescription. 

\end{itemize}

\subsection{Constraints on instanton shape}
\label{sec:fix}

\begin{figure}[tbp]
\centering 
\underline{$\gamma_0=4$}\hspace{6cm} \underline{$\gamma_0=6.2$} \hspace{-1cm}

\vspace{0.4cm}
\hbox{\includegraphics[width=.5\textwidth]{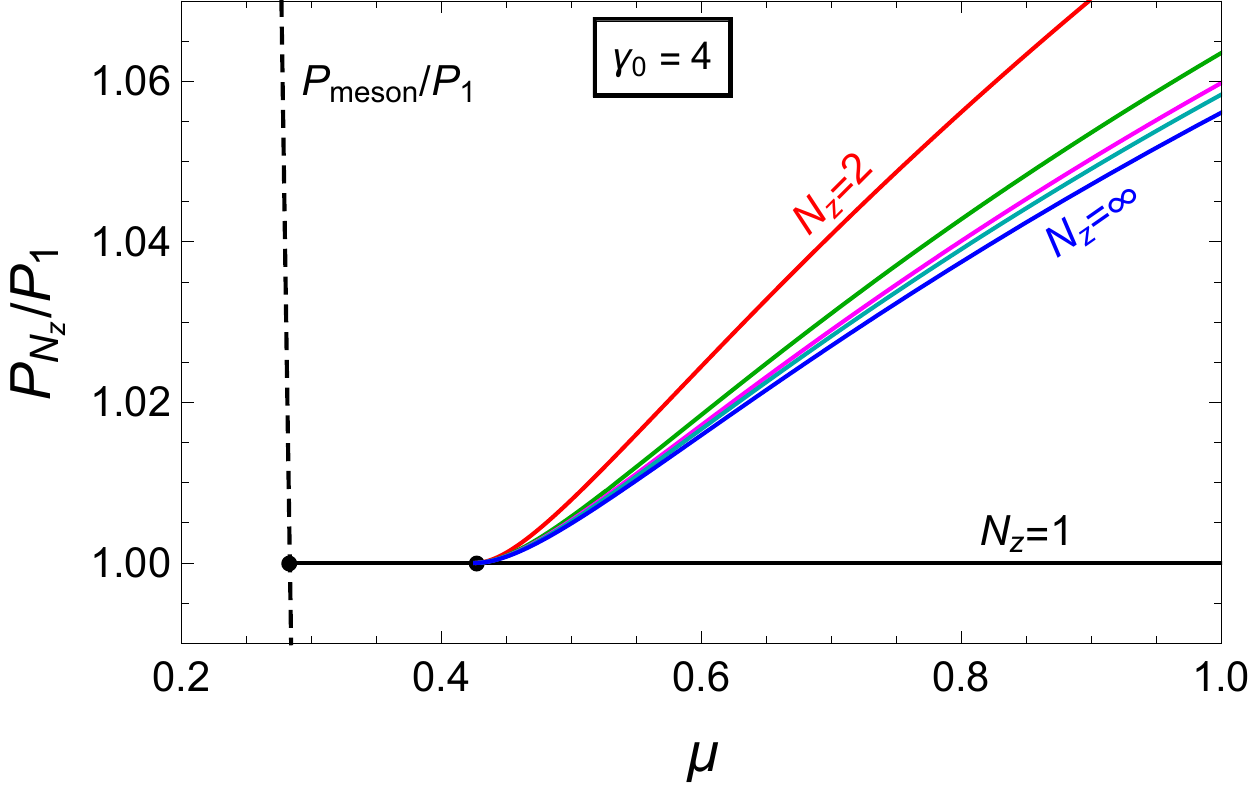}\includegraphics[width=.5\textwidth]{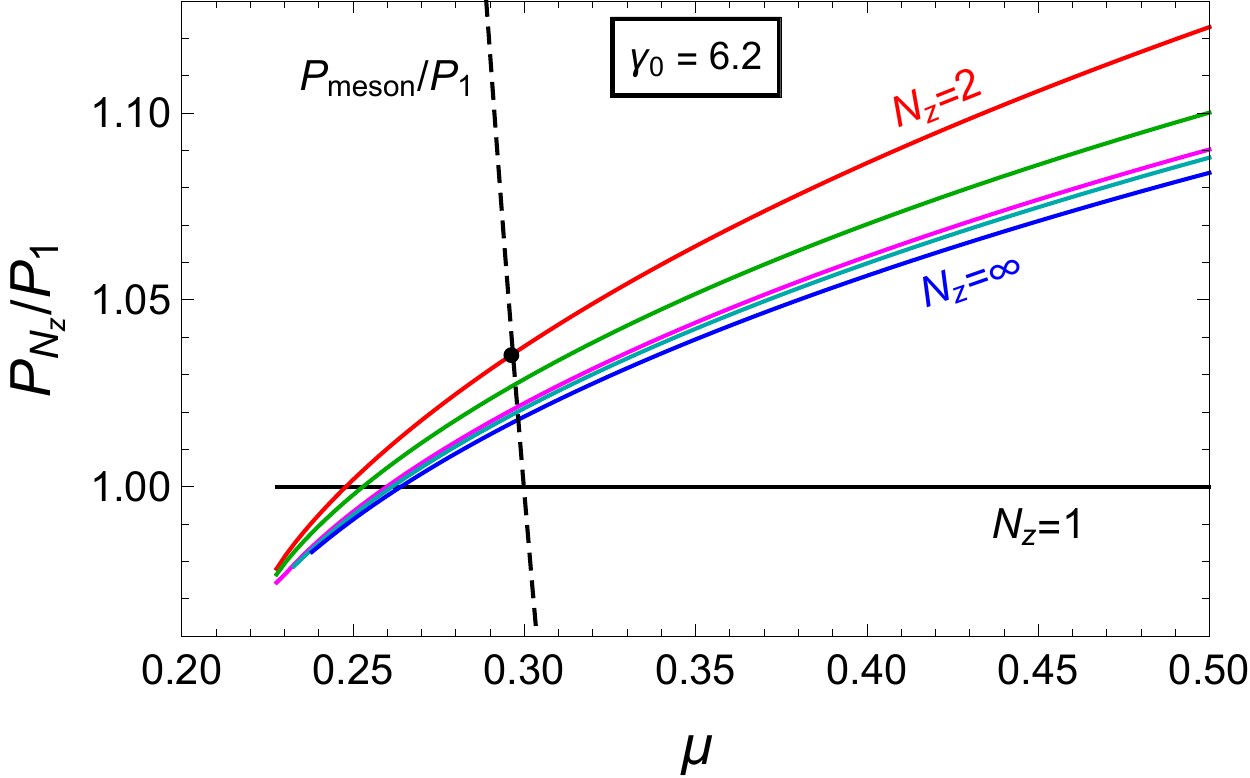}}

\hspace{0.19cm}\hbox{\includegraphics[width=.485\textwidth]{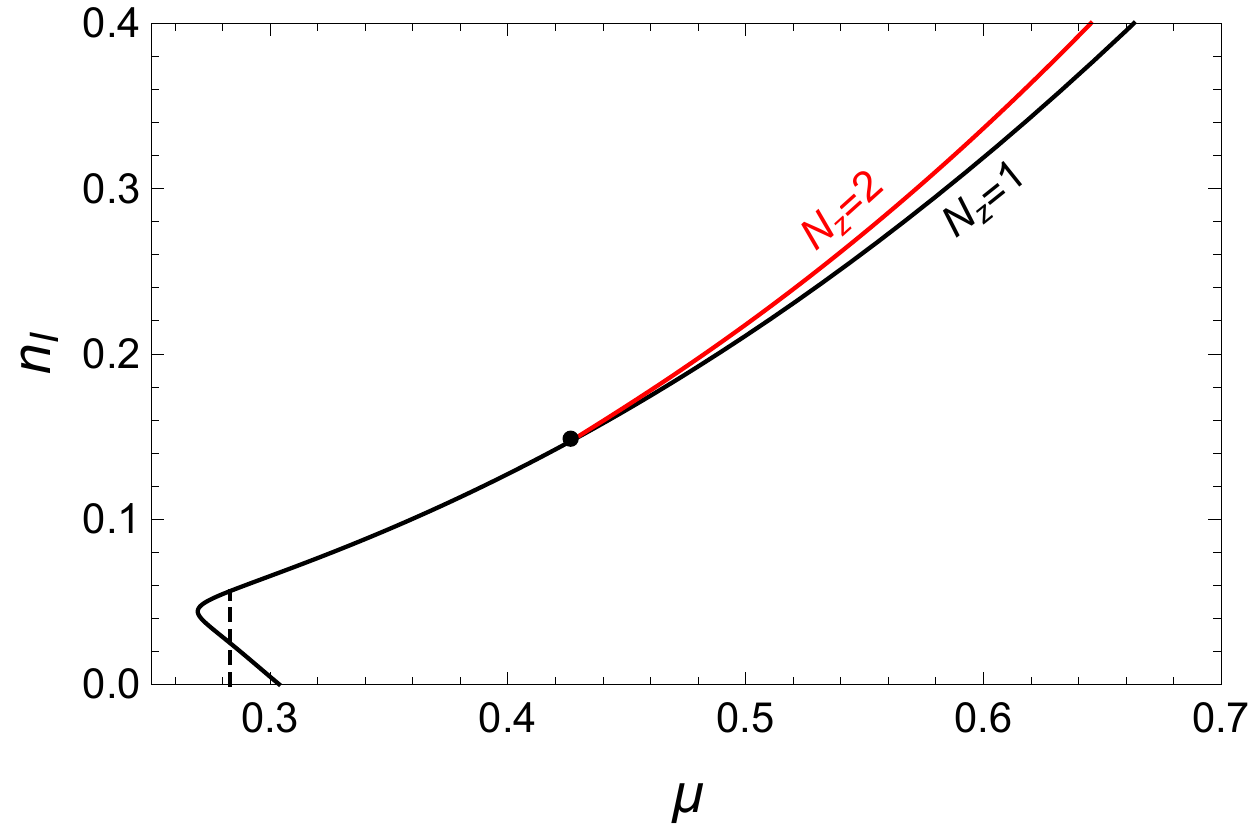}\hspace{0.1cm}\includegraphics[width=.495\textwidth]{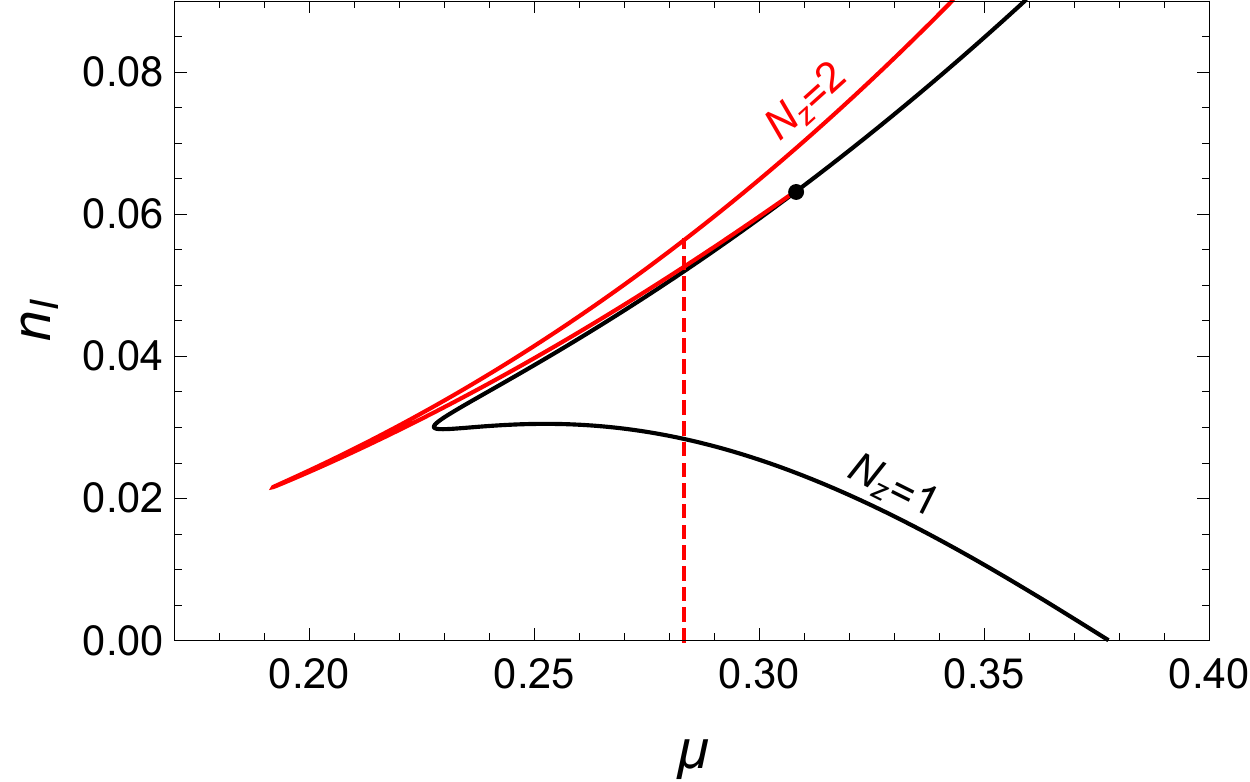}}
\caption{\label{fig:Om123} 
Pressures and corresponding densities for an instanton width $\rho_0=2.5$ and two different instanton deformations, $\gamma_0=4$ (left column) and $\gamma_0=6.2$ (right column). Upper row:
comparison of pressures of baryonic matter $P_{N_z}$, with $N_z=1,2,3,4,5,\infty$ instanton layers in the bulk (solid lines) and the mesonic phase $P_{\rm meson}$ (dashed lines). In both panels, there is a 
first order phase transition from the mesonic phase to baryonic matter. In the left panel, this transition is to the $N_z=1$ phase, and the solutions for $N_z>1$ only start to exist at a larger chemical potential. In the 
right panel, the solutions $N_z>1$ start to exist below the baryon onset, and the transition is to the $N_z=2$ phase. There are energetically disfavored branches whose pressure we have not shown
(for $N_z=1$ in the left panel and for all $N_z$ in the right panel). Lower row: corresponding baryon densities for $N_z=1$ and $N_z=2$, showing all solutions, including the energetically disfavored branches. The dashed
vertical lines indicate the baryon onset. The curves for $N_z=3,4,5,\infty$ are not shown since they would be difficult to distinguish from the $N_z=2$ curve on the given scale.} 
\end{figure}

We now turn to approach (ii), where we impose the constraints (\ref{rhogamma}) on $\rho$ and $\gamma$. One of the crucial differences to the previous section 
is that now we do find solutions for all $N_z$. We thus have to compare the free energies of all phases with different numbers of instanton layers in the regime where their solutions coexist. 

Let us start with a specific choice of parameters, $\rho_0=2.5$, $\gamma_0=4$. 
The results for the free energies and corresponding densities are shown in the left column of Fig.\ \ref{fig:Om123}. [The ratio of the free energy densities of two phases is obviously the same as the ratio of pressures because the minus signs 
in the free energies simply cancel. 
However, since a ratio {\it larger} than 1 (and a {\it larger} pressure, but a {\it lower} free energy) is favorable, it is somewhat more natural to label the vertical axis with the ratio of pressures, 
as in Fig.\ \ref{fig:nIuc}, although in the text we continue to speak of the free energy of the phases.] 
The first important result is that the solution for $N_z=1$ is multi-valued in a certain regime of chemical potentials and as a consequence there is a first-order baryon onset, in contrast to the result of the previous section. 
In the left column, the solutions for all $N_z>1$ are single-valued and start to exist at the same point. As soon as they exist, their free energy is lower (and their density larger) than that of the $N_z=1$ solution. We find that the free energy of the $N_z=2$ solution is lowest, and $\Omega_{N_z=2}<\Omega_{N_z=3}<\ldots <\Omega_{N_z=\infty}<\Omega_{N_z=1}$. 
Therefore, there is a transition from the baryonic phase with a single instanton layer to a baryonic phase where two instanton layers separate in the bulk. This transition is smooth.
The right column of Fig.\ \ref{fig:Om123} shows the result for the same $\rho_0$, but a larger deformation parameter, $\gamma_0=6.2$. Now, the solutions for {\it all} $N_z$ are multivalued, and there is a 
first-order baryon onset directly to the phase with $N_z=2$. In both cases, we have thus, by imposing the constraint (\ref{rhogamma}), arrived at a first-order phase transition to baryonic matter, as expected 
from real-world nuclear matter. 

\begin{figure}[tbp]
\centering 
\underline{$\mu=1$}\hspace{4cm} \underline{$\mu=3$}\hspace{4cm}\underline{$\mu=6$}

\vspace{0.4cm}
\hbox{\includegraphics[width=.33\textwidth]{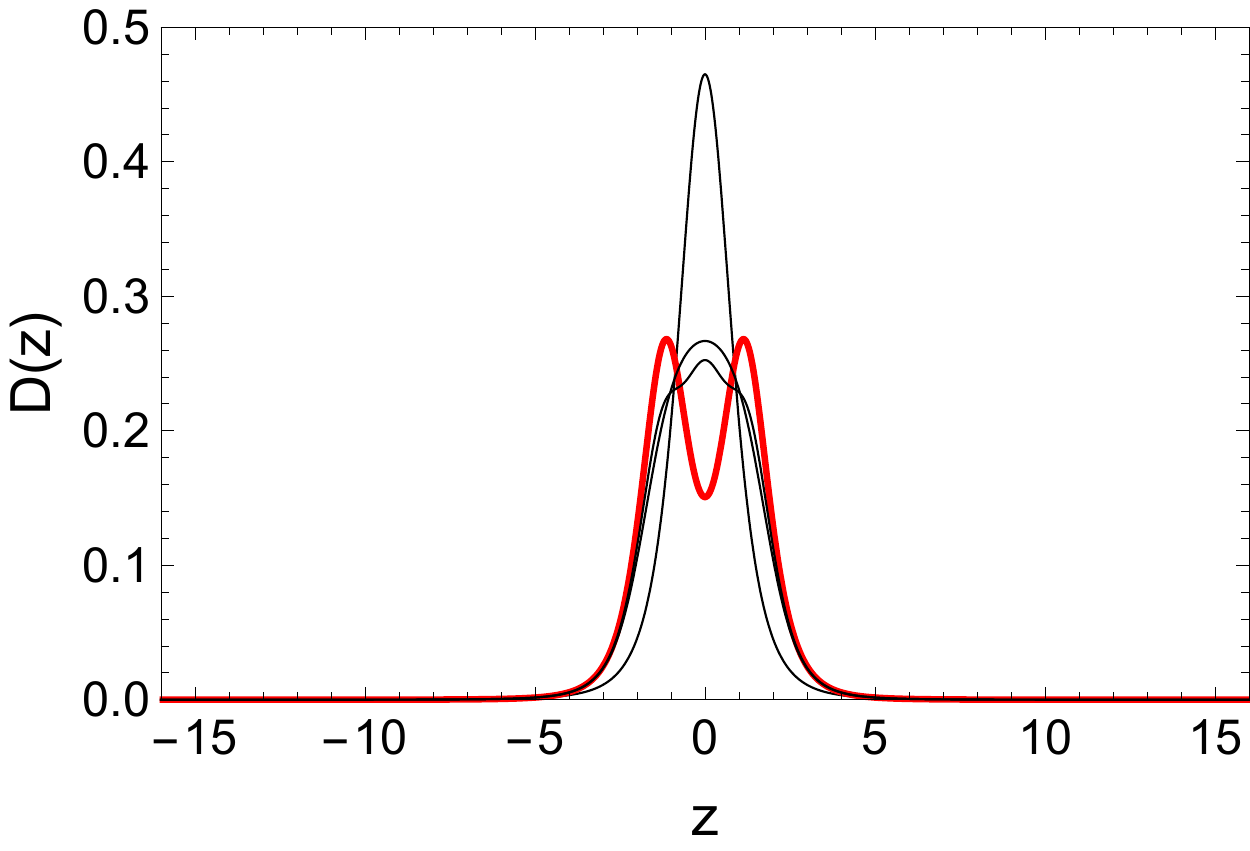}\includegraphics[width=.33\textwidth]{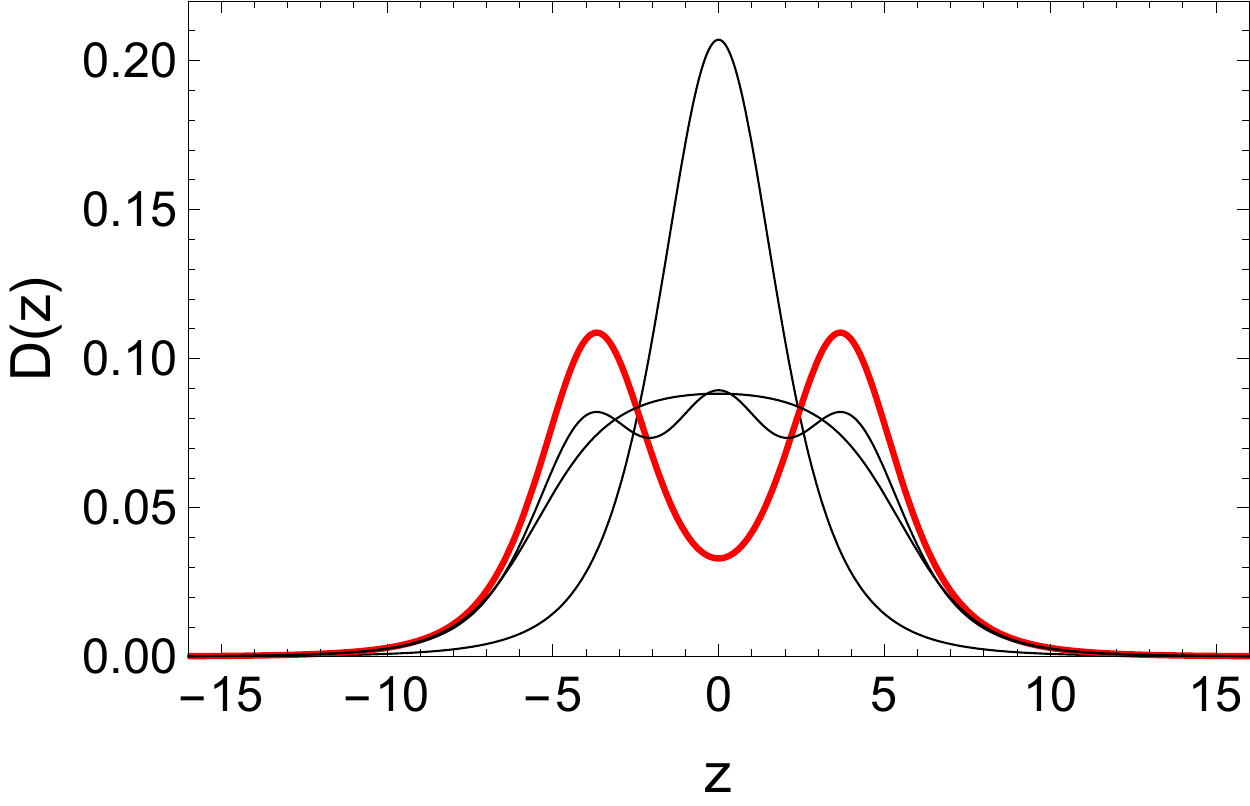}\includegraphics[width=.33\textwidth]{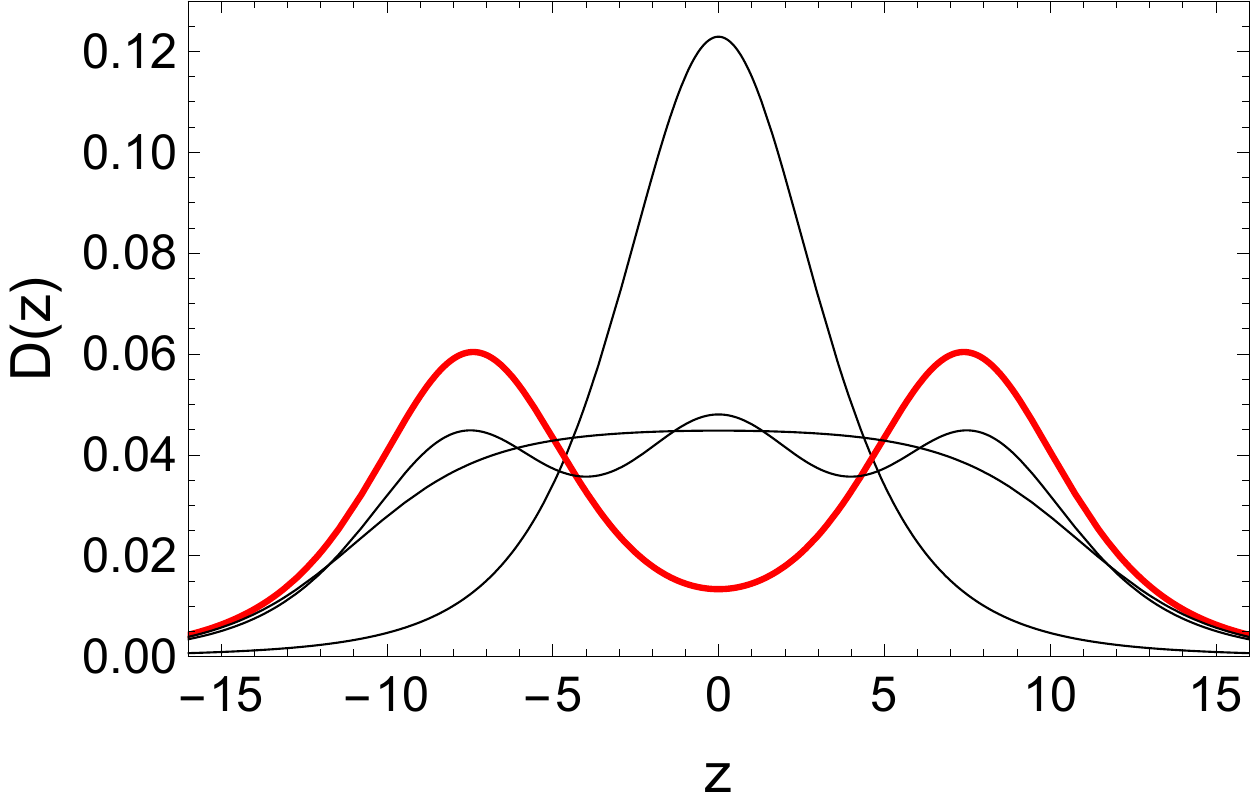}}

\vspace{0.5cm}
\hbox{\includegraphics[width=.33\textwidth]{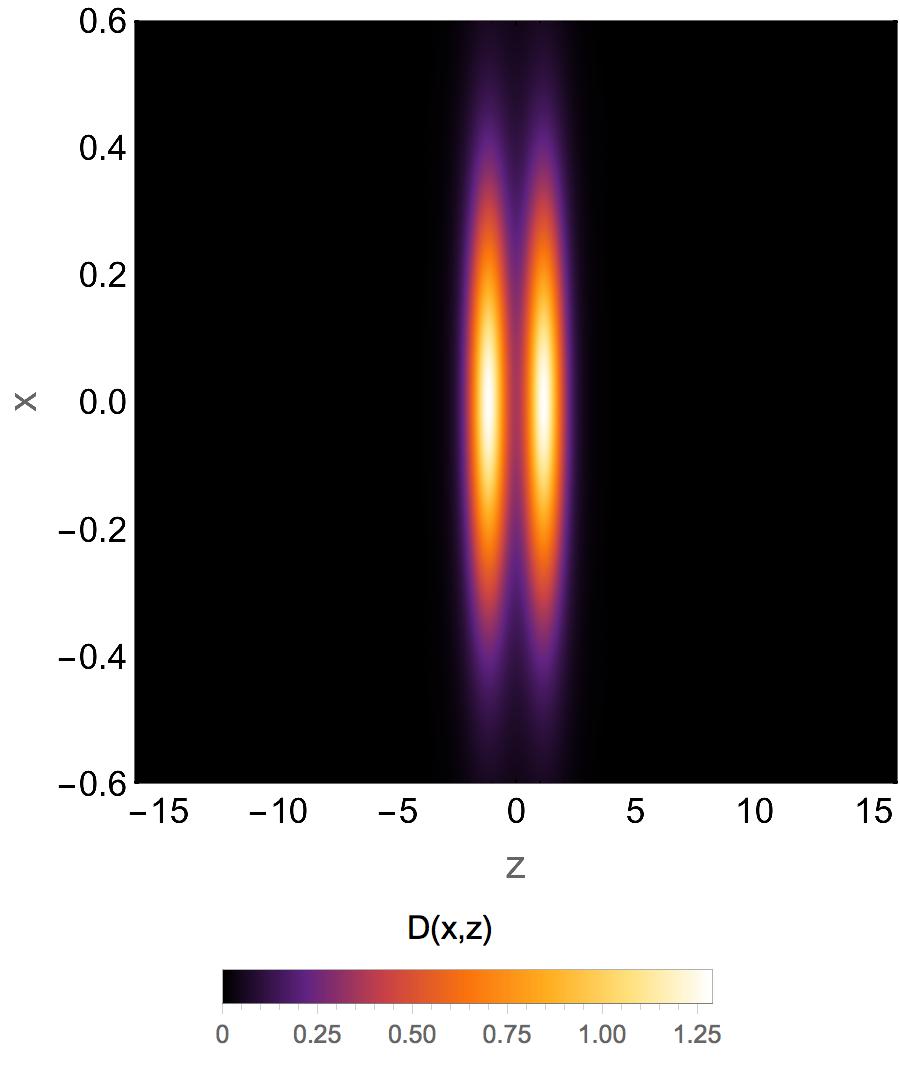}\includegraphics[width=.33\textwidth]{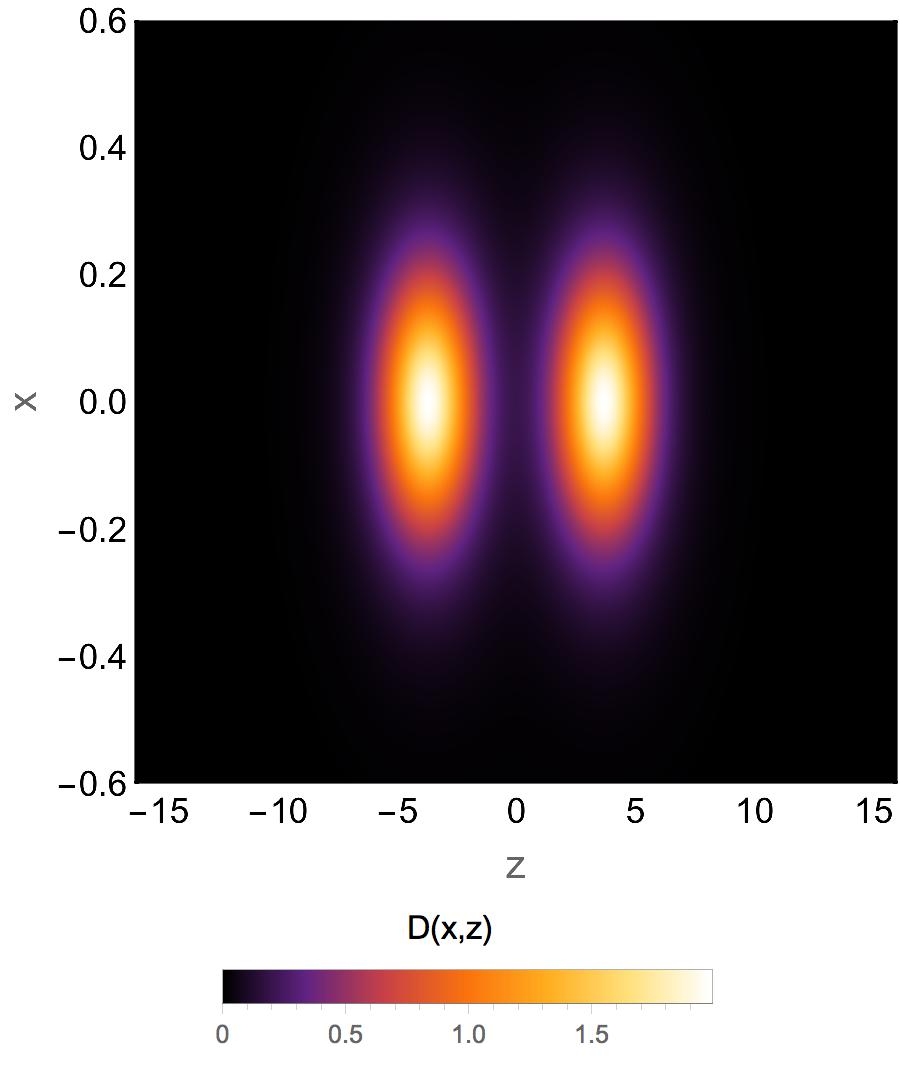}\includegraphics[width=.33\textwidth]{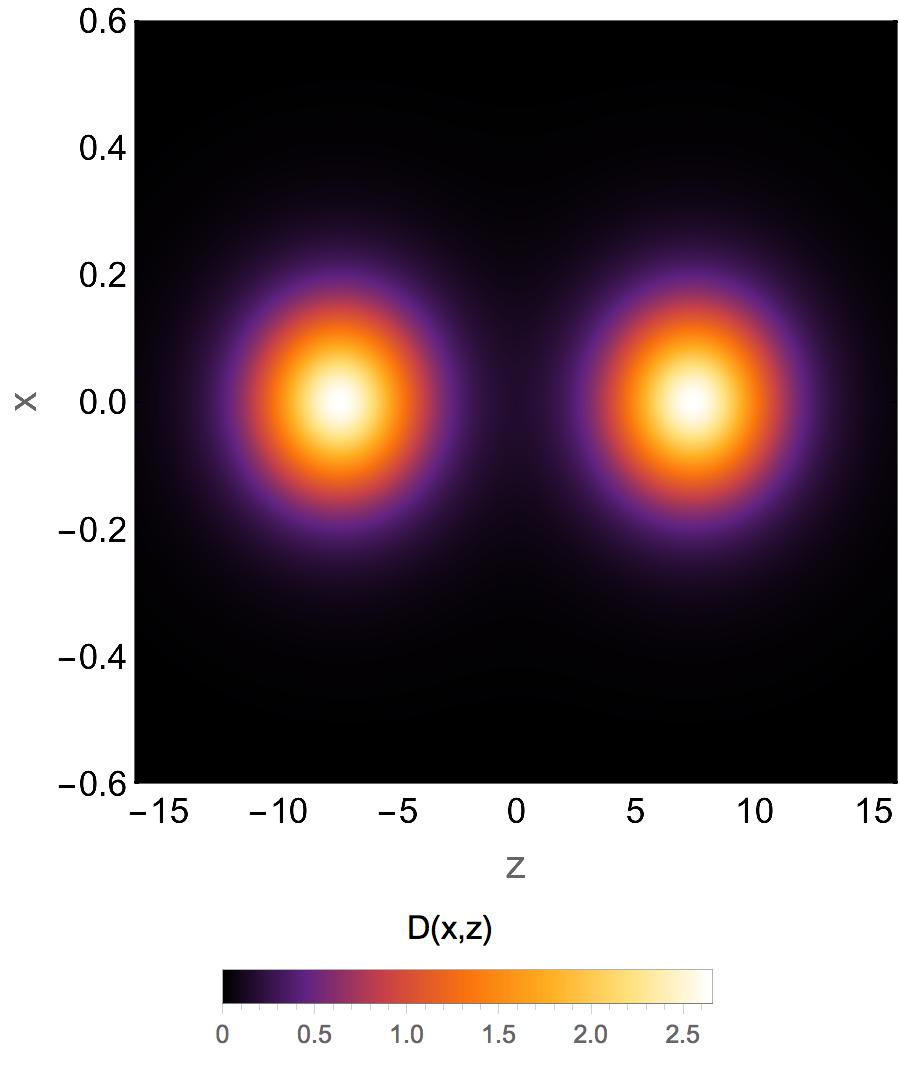}}

\vspace{0.5cm}
\hbox{\includegraphics[width=.33\textwidth]{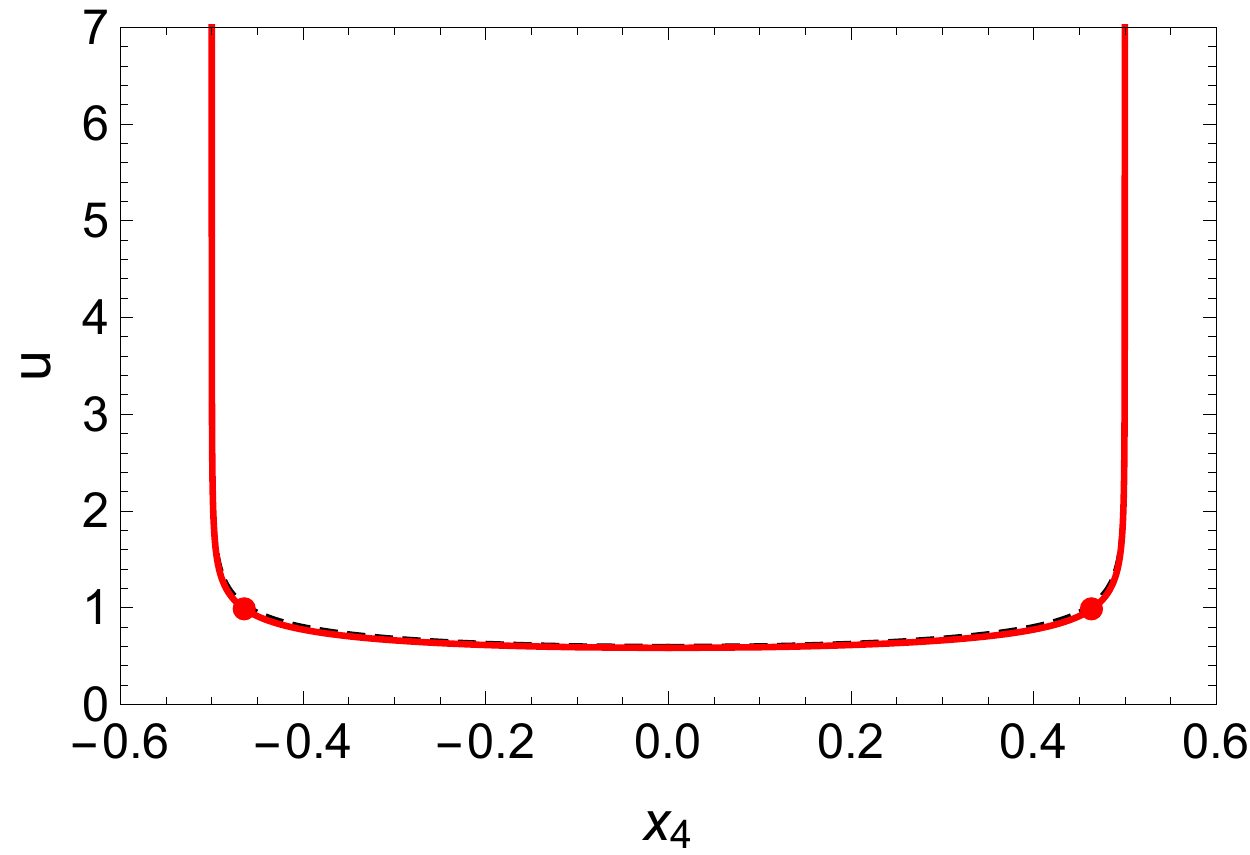}\includegraphics[width=.33\textwidth]{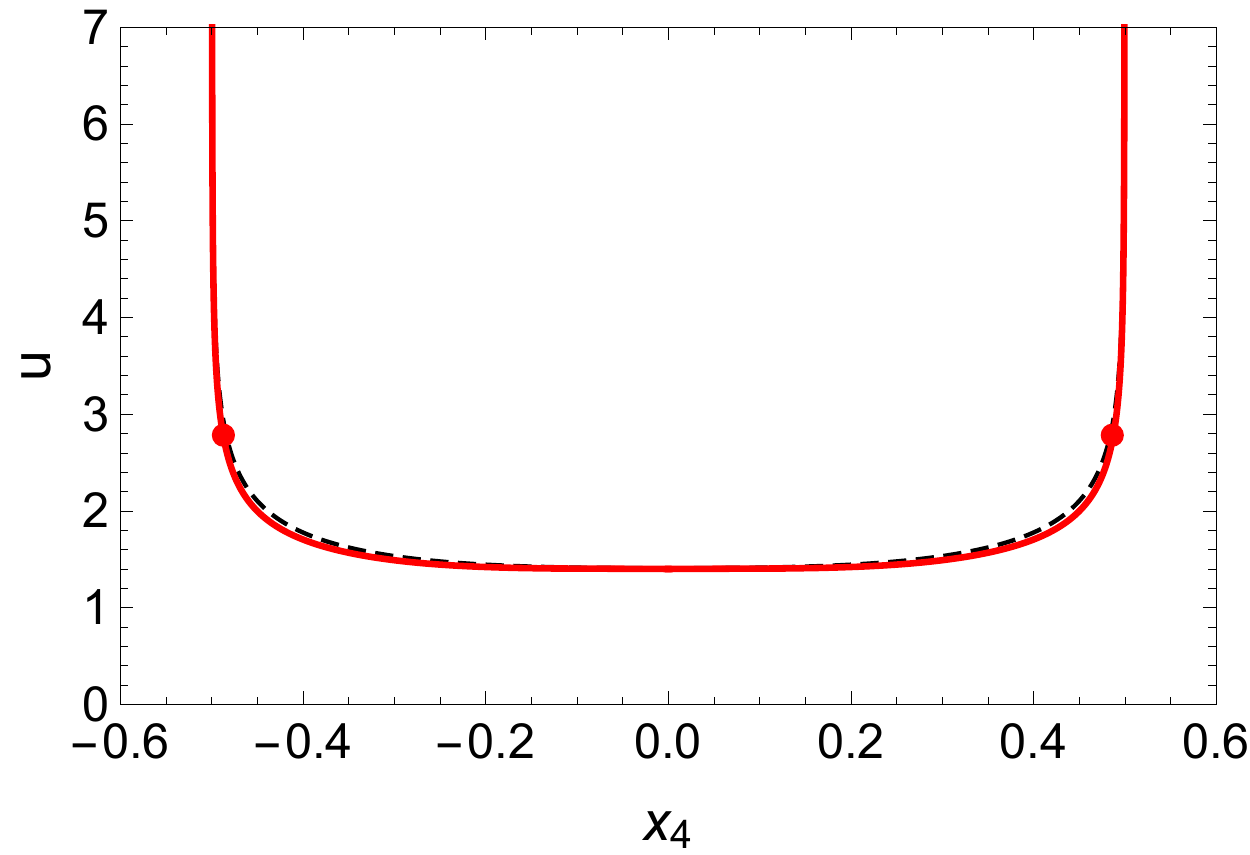}\includegraphics[width=.33\textwidth]{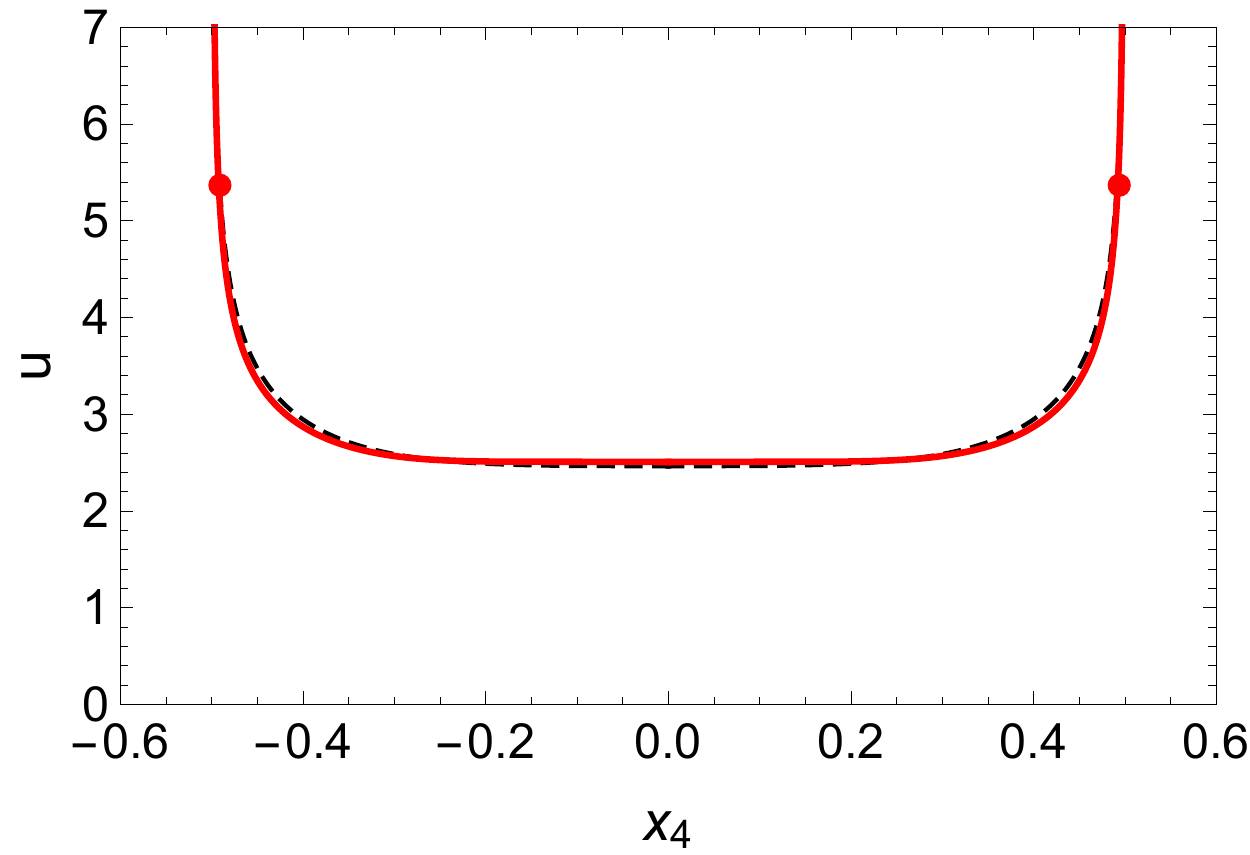}}
\caption{\label{fig:bumps} Instanton profiles (first two rows) and corresponding embedding functions of the flavor branes (third row) for $\rho_0=2.5$, $\gamma_0=4$ and three different chemical potentials, $\mu=1$ (left column),
$\mu=3$ (middle column), $\mu=6$ (right column). 
First row: profiles $D(z)$ in the holographic direction $z$. For chemical potentials above 
a certain critical potential ($\mu\simeq0.43$ for the given parameters, see left column of Fig.\ \ref{fig:Om123}) solutions for all $N_z$ exist, but the energetically preferred solution is $N_z=2$ [thick (red) line]. 
We have also plotted the energetically disfavored solutions $N_z=1,3,\infty$ [thin (black) lines]. 
Second row: energetically preferred profiles $D(x,z)$ in the holographic and radial directions, showing that the instanton gets elongated along the $z$ direction for large densities.  
Note that the vertical axis in the first row and the color scale in the second row is adjusted for each panel (whereas the 
$z$ and $x$ intervals are fixed). Third row: the solid (red) curve is the embedding of the preferred $N_z=2$ solution, with the 
dots marking the centers of the two instanton layers (this is the calculated version of the cartoon in Fig.\ \ref{fig:cylinders}). 
The dashed (black) curve, which is barely distinguishable from the solid curve, is the embedding for the (energetically disfavored) $N_z=\infty$ solution. 
}
\end{figure}

In Fig.\ \ref{fig:bumps} we show the details of the solution obtained for the parameters from the left column of Fig.\ \ref{fig:Om123}. In the upper row, the instanton profile 
in the holographic direction $D(z)$, see Eq.\ (\ref{Dz}), is plotted for three different chemical potentials. The chemical potentials chosen here are all above the onset of the $N_z=2$ solution, i.e., the (red) thick line that represents the solution with lowest free energy 
 has always two  maxima, symmetrically placed around $z=0$. These maxima move apart with increasing chemical potential. The spreading of the instantons in the bulk with 
 increasing density was observed previously in the Sakai-Sugimoto model and related models in various different approximations. Firstly, it was suggested in the confined phase of the Sakai-Sugimoto model within a simple approximation unrelated to any single-instanton solution \cite{Rozali:2007rx}. 
A similar observation, taking into account a crystalline structure in both spatial and holographic coordinates, led to the 
term "baryonic popcorn" \cite{Kaplunovsky:2012gb,Kaplunovsky:2015zsa}, referring to a successively increasing number of instanton layers in the holographic direction with increasing density -- in contrast to our approximation,
where at most two layers are favored. The observation of baryonic popcorn was confirmed in a full numerical calculation within a simpler, 2+1 dimensional 
model \cite{Bolognesi:2013jba,Elliot-Ripley:2015cma}. These results in the literature suggest that the occurrence of multiple instanton layers, or, more generally, the spreading of the instantons away from the tip of the connected flavor branes, at large baryon density is a general feature, and it is intriguing that our simple approximation for homogeneous baryonic matter, based on the flat-space BPST instanton solution, shows the same feature, if we enforce the constraints (\ref{rhogamma}). 

In the second row of Fig.\ \ref{fig:bumps} we show the same instanton profiles, but now in the two-dimensional space of holographic and spatial directions. We recall that even though 
we have averaged over position space before solving the equations of motion, we can still go back to the instanton profile $D(x,z)$ from Eq.\ (\ref{Dxz}) and ask how this profile looks for different chemical potentials. 
Although we have fixed $\rho_0$ and $\gamma_0$, the instanton width $\rho$ and deformation $\gamma$ remain nontrivial functions of the chemical potential due to their dependence on $u_c$, 
which is determined dynamically. 
The figure shows that the instantons not only develop a second layer at large $\mu$, but also become elongated in the holographic direction: with increasing density, the instantons get wider in the holographic direction because 
$\rho \propto u_c$ increases (already obvious from the first row of the figure), and narrower in the spatial direction because $\rho/\gamma \propto u_c^{-1/2}$ decreases. 

The third row of Fig.\ \ref{fig:bumps} shows the embedding of the flavor branes in the background geometry for the same chemical potentials as the first two rows. These plots illustrate the instanton layers in the 
subspace spanned by $u$ and $x_4$. They show in particular that, due to the flatness of the brane profile just above $u_c$, a small distance between the instanton layers in the $z$ (or $u$) coordinate can 
result in a large distance along the brane profile, if the instantons are close to $u_c$. We also see that the shape of the profile does not change qualitatively with density (apart from moving up), even though the 
instantons move from the flat part of the profile up to the almost vertical segments of the branes. It does not change much with the number of instanton layers either, which can be seen from the comparison with the 
$N_z=\infty$ embedding (we have checked that the embeddings for other values of $N_z$ are also barely distinguishable from the shown curves).  In other words, the flavor branes do not seem to care much about 
how many instanton layers they carry and where they sit. 

\begin{figure}[tbp]
\centering 
\hbox{\includegraphics[width=.5\textwidth]{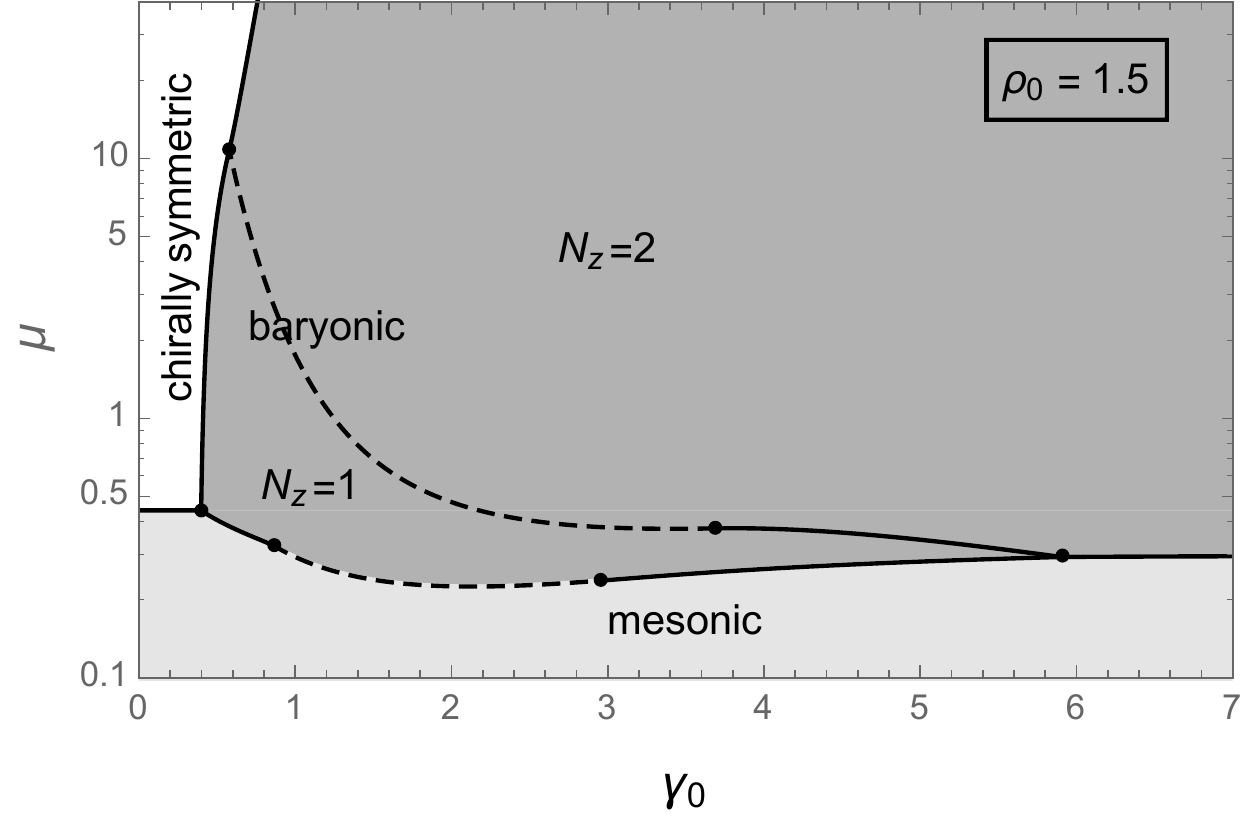}\includegraphics[width=.5\textwidth]{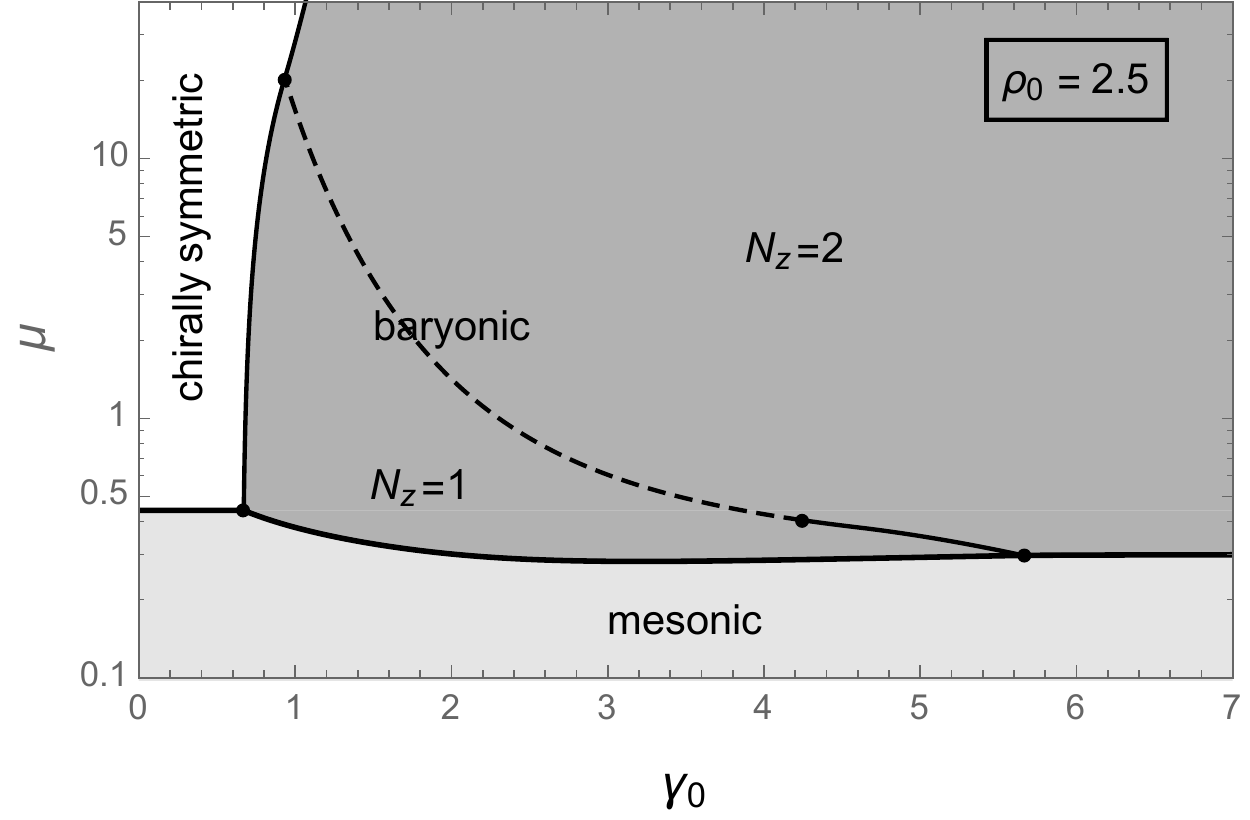}}
\caption{\label{fig:phases} Phase diagrams with constraints on instanton width and deformation [approach (ii)]. Solid and dashed lines are first and second order phase transition lines, respectively. There are two different 
baryonic phases, with one, $N_z=1$, or with two, $N_z=2$, instanton layers.}
\end{figure}

As we have seen in Fig.\ \ref{fig:Om123}, different choices of $\rho_0$ and $\gamma_0$ can lead to qualitatively different behaviors of the system. In principle, we can now scan the entire two-dimensional parameter space, and determine the phases and phase transitions for all chemical potentials. Keeping $T=0$, this would result in a three-dimensional $\rho_0$-$\gamma_0$-$\mu$ phase diagram. We present two two-dimensional slices of this phase
diagram in Fig.\ \ref{fig:phases}, where we have fixed $\rho_0$ and computed all phase transition lines in the $\gamma_0$-$\mu$ plane. In the right panel, $\rho_0=2.5$, i.e., the results of  
Figs.\ \ref{fig:Om123} and \ref{fig:bumps} are obtained along two vertical lines in that panel. The left panel has been calculated with a smaller value, $\rho_0=1.5$. In appendix \ref{app:phases} we explain how we have 
computed the various phase transition lines and critical points. 

Before we come to the observations, let us add a remark regarding the connection of the phase diagrams to the results of the previous section. 
The minimization carried out in Sec.\ \ref{sec:tseytlin} determines a trajectory of the system through the three-dimensional space spanned by $\mu,\rho_0,\gamma_0$. This trajectory intersects each of the slices shown in Fig.\ \ref{fig:phases} in a point. We find that for both slices this point lies in the region where $N_z=2$ (for the left slice this can be read off of the upper right panel of Fig.\ \ref{fig:nIuc}). 
One might thus
naively conclude that we have found a stationary point of the free energy with $N_z>1$, even though we have argued in Sec.\ \ref{sec:tseytlin} that such a point does not exist. But this conclusion is not correct: 
there may very well be a minimum at $N_z=2$ under the constraint
of a fixed pair ($\rho_0,\gamma_0$), but no minimum for the same $\mu$ with $N_z=2$ if we search for the minimum in the entire parameter space, including $\rho_0$ and $\gamma_0$. 

The main observations of Fig.\ \ref{fig:phases} are as follows. We see that the baryon onset can be of first order, as in Fig.\ \ref{fig:Om123}, but also of second order, as in Sec.\ \ref{sec:tseytlin}. It appears that 
smaller values of $\rho_0$ can produce a second order onset. This makes sense because by decreasing $\rho_0$ the instanton width becomes smaller, i.e., we approach the pointlike limit, and we 
know that the pointlike approximation predicts a second order onset. Small values of $\rho_0$ (like high densities and large values of $\gamma_0$) also seem to prefer instanton spreading. This suggest that, if we were to 
extend the pointlike approximation by allowing for instanton repulsion, we would presumably find the degenerate, delta-peaked instantons move up in the holographic direction with increasing chemical potential. 
This calculation might be of some interest because it would be the simplest system in which the instanton repulsion could be observed, possibly allowing for some analytic results, at least in certain limits such as large densities. On the other hand, our present results show that the pointlike approximation is not a good approximation for large densities and thus we do not include this calculation here. 
In the present scenario, the instanton width $\rho=\rho_0 u_c$ is always nonzero because $u_c$ never goes to zero. Thus, one way of thinking about the results is that 
in approach (i) of Sec.\ \ref{sec:tseytlin}, the system chooses to have pointlike baryons at infinitesimally small densities, resulting in a second order onset, and here, in approach (ii), we forbid pointlike 
baryons via the external constraint $\rho_0>0$ and in accordance with expectations from finite-$\lambda$ corrections, and thus we are able to see a first order onset. 

For sufficiently small values of $\gamma_0$, i.e., stretching the instanton in the spatial direction compared to the holographic direction, there is no baryonic phase at all: the vacuum is directly superseded by quark matter. For slightly larger values of $\gamma_0$ the baryon onset is followed by a transition to chirally restored matter (upon increasing $\mu$ at fixed $\rho_0$ and $\gamma_0$). The numerical results suggest that there is no chiral transition at all for sufficiently large 
$\gamma_0$. However, the numerics become difficult at very large $\mu$, and thus we cannot say this with certainty. In any case, except for a very narrow regime, 
the chiral phase transition occurs -- if at all -- at {\it much} larger values of the chemical potential than the baryon onset (note the logarithmic $\mu$ scale in Fig.\ \ref{fig:phases}).

It is interesting to note that the overall structure of the phase diagrams is very similar to the one obtained with the "homogeneous ansatz",  with $\gamma_0$ replaced by the 't Hooft 
coupling $\lambda$, see Fig.\ 7 of Ref.\ \cite{Li:2015uea} (by comparing the equations it is obvious that the instanton deformation $\gamma$ in the present instantonic ansatz plays a very similar role to $\lambda$ in the homogeneous ansatz). One difference is that the 
chiral phase transition line bends in the other direction: within the homogeneous ansatz the baryon onset is never followed by a chiral phase transition if $\lambda$ is held fixed. In both cases, $\mu$-dependent 
parameters [here $\rho_0(\mu), \gamma_0(\mu)$, there $\lambda(\mu)$] would allow for an equation of state that shows a first-order baryon onset {\it and} a transition to quark matter at moderately large densities. 
From a purely phenomenological point of view, it might be tempting to search for such a suitable $\mu$ dependence: one might add $\rho_0$ and $\gamma_0$ 
to the three free parameters of the model and fit them to known properties of nuclear matter at the saturation density, or to the (poorly known) critical chemical potential of the chiral phase transition. 
However, fitting them in a density-dependent way would necessarily include some arbitrariness or, at best, some 
extrapolation to large densities. And, from a theoretical point of view, there is no reason for using the phase diagrams of Fig.\ \ref{fig:phases} and moving through them with externally
given functions $\rho_0(\mu)$, $\gamma_0(\mu)$. We do know how $\rho$ and $\gamma$ "want" to behave as a function of $\mu$ in the given approximation. This was discussed in Sec.\ \ref{sec:tseytlin} and has led to unphysical results, a second-order baryon onset and no chiral restoration. Therefore, the results of the present section should be understood as a step towards a better understanding of the instanton approach to baryonic matter, its relation to the homogeneous ansatz and, in future 
work, towards further improvement of the approximation, rather than a straightforward recipe for constructing a strong-coupling equation of state for dense matter inside compact stars.

\section{Summary and outlook}
\label{sec:summary}

We have investigated homogeneous baryonic matter at zero temperature in the decompactified limit of the Sakai-Sugimoto model. Our main point was to improve existing approximations based on flat-space instantons
and to ask whether these improvements bring us closer to real-world nuclear matter, in particular whether they give rise to a first-order baryon onset and a chiral transition to quark matter at high densities. 
Motivated by results of holographic baryons in the vacuum, we 
have introduced a deformation parameter into the instanton ansatz that allows for an anisotropy in the space of holographic and spatial directions. While in the spatial direction we have employed an averaging procedure, accounting for homogeneous matter in a very simple way, we have introduced instanton repulsion in the bulk: the instantons are allowed to spread out in the holographic direction 
in the form of a number of  instanton layers, this number and the distance between the layers being determined dynamically. 

We have found that if we minimize the free energy with respect to the width and the deformation of the instantons (and with respect to the various other parameters of our ansatz) that (1) there is a (unphysical)
second-order baryon onset, and at the onset the instantons are pointlike, reproducing the approximation of degenerate, delta-peaked instantons, (2) baryons (unphysically) refuse to go away at large densities, i.e., 
there is no chiral restoration, (3) at large densities, the instantons tend to get elongated along the holographic direction, and (4) the instantons prefer to sit all at the same point in the bulk, i.e., the number of instanton layers remains 1 for all densities.

Besides this most straightforward approach we have also worked with external constraints on the width and the deformation of single instantons  -- being aware that our simple 
approximation cannot capture the shape of the full solution -- and studied their effect on the many-instanton system. More precisely, we have constrained the width $\rho$ to be proportional to the  (density-dependent) 
location of the tip of the connected flavor branes $u_c$ and the deformation parameter $\gamma$ to be proportional to $u_c^{3/2}$, and treated the proportionality constants $\rho_0$ and $\gamma_0$ as free parameters.
Interestingly, with these constraints, which in particular result in non-pointlike instantons at all densities, we do find that at sufficiently large densities the instantons 
are divided into two layers. With increasing density these layers move up in the holographic direction, away from the 
tip of the connected flavor branes. All higher numbers of layers turn out to be energetically disfavored. In particular, we have included the possibility of infinitely many layers, corresponding to a smeared distribution along the 
holographic direction. Our results regarding the instanton layers is interesting in view of various other approximations and approaches in the same model, and complete solutions in simplified models, which show similar 
effects \cite{Rozali:2007rx,Kaplunovsky:2012gb,Kaplunovsky:2015zsa,Bolognesi:2013jba,Elliot-Ripley:2015cma,Gorsky:2015pra}. We have also found that in the 
presence of the constraints on width and deformation, we do see a first-order baryon onset and chiral restoration at large densities, as expected from QCD: for sufficiently large $\rho_0$, the 
baryon onset is first order, and for a small range of values of $\gamma_0$, baryonic matter is superseded by chirally restored quark matter, with the critical chemical potential for chiral restoration 
being very sensitive on $\gamma_0$. 

In conclusion, we have shown that the present instanton approximation -- if evaluated at its stationary point -- is too simplistic to show multiple instanton layers (which are suggested by other approximations of the model), 
and it does neither show a first-order baryon onset nor chiral restoration (which occur in the real world). We have shown that by imposing external constraints on the shape of the single instantons, 
these features do appear. 

For a short discussion of future perspectives we first recall that this work has been, to a large extent, motivated by a phenomenological question: can we come up with a strong-coupling model description of dense nuclear and 
quark matter within a single model? In principle, we could use our results to find such a model description. However, to fulfill even the most fundamental requirements of real-world matter, this would require 
a theoretically ill-motivated, density-dependent choice of $\rho_0$ and $\gamma_0$, which renders any resulting predictions questionable. Therefore, the results should mainly be considered as a further theoretical step towards such a model. We have embedded our ansatz into a more general setup, which allows for systematic improvements. For instance, it would be very useful to extend our ansatz to one that incorporates a nontrivial dependence on the 't Hooft coupling
$\lambda$, even though a systematic treatment of finite-$\lambda$ effects would require string corrections, which is very difficult.  
It would also be interesting to gain a deeper understanding of the relation between our instanton approach and the results of the "homogeneous ansatz" \cite{Rozali:2007rx,Li:2015uea} 
and of the relation of both to the full solution. For the latter it is useful to retreat to simpler models or simplifications of the Sakai-Sugimoto model, where the full solution is available and can be compared to the 
various approximations \cite{Bolognesi:2013jba,Elliot-Ripley:2015cma}. Other promising extensions, within the present ansatz or one that is further improved, are to include nonzero temperatures (all our equations contain the full temperature dependence, but we have restricted ourselves to zero temperature in the numerical results), to include an isospin chemical potential, the possibility of a chiral density wave, or to add an external magnetic field.

\acknowledgments

We would like to thank Matthew Elliot-Ripley, Anton Rebhan, and Paul Sutcliffe for helpful discussions and acknowledge support from the {\mbox NewCompStar} network, COST Action MP1304. 
F.P.\ is supported by the Austrian Science Fund (FWF) under project no.\ P26328-N27, and A.S.\ is supported by the Science \& Technology Facilities Council (STFC) in the form of an Ernest Rutherford Fellowship.

\appendix 

\section{Instantons in the vacuum from the Yang-Mills approximation}
\label{app:YM}

In this appendix we discuss the single-instanton solution in the deconfined geometry. The calculation differs from that in the main part in several aspects: besides considering a single instanton, 
and not a many-instanton system, 
we use the Yang-Mills (YM) action, and we employ an expansion for large $\lambda$, in particular using the leading order result in $\lambda$ for the embedding function for the flavor branes. As a consequence, we are able to solve
the equations of motion for all gauge fields analytically, including the dependence on the position space coordinates. The analogous calculation in the confined geometry can be found in the literature, for maximally separated 
flavor branes \cite{Hata:2007mb} and with a general, not necessarily maximal, separation \cite{Seki:2008mu}. The present calculation in the deconfined geometry in particular yields a nontrivial temperature dependence for 
the instanton width and the instanton deformation. 

In the main part, we have ignored any dependence of the abelian gauge field $\hat{A}_0$ on $\vec{X}$, and thus we first have to reinstate $\hat{F}_{0i}$ into Eq.\ (\ref{DBIdeconf}) [or Eq.\ (\ref{factorized}), 
which only differs from Eq.\ (\ref{DBIdeconf}) by  
terms of order $F^4$, which shall be neglected in this appendix] and insert the result into the DBI action (\ref{SDBI}). Then, we derive the YM action by expanding in the field strengths up to order $F^2$, which 
results in the action  
\bea
S &\simeq& S_0 + S_{\rm YM} + S_{\rm CS} \, , 
\eea
where 
\bea
S_0 &=& \frac{\lambda N_cN_fM_{\rm KK}^4}{24\pi^3}\frac{V}{T}\lambda_0^2 \int_{u_c}^\infty du\, u^{5/2}\sqrt{1+u^3f_Tx_4'^2} \, ,
\eea 
with $\lambda_0\equiv \lambda/(4\pi)$, is a purely geometric term, not depending on any gauge fields, and where 
\begin{subequations}
\bea
S_{\rm YM} &=& \frac{\lambda N_cM_{\rm KK}}{48\pi^3T} \int d^3 x \int_{u_c}^\infty du\, u^{5/2}\sqrt{1+u^3f_Tx_4'^2} \non[2ex]
&&\times\left(\frac{2\lambda_0^2\Tr[\hat{F}_{0i}^2]+f_T\Tr[F_{ij}^2]}{2f_Tu^3}
+\frac{\lambda_0^2\Tr[\hat{F}_{0u}]^2+f_T\Tr[F_{iu}^2]}{1+u^3f_Tx_4'^2}\right) \, , \\[2ex]
S_{\rm CS} &=& \frac{\lambda N_c M_{\rm KK}}{32\pi^3 T} \int d^3 x \int_{-\infty}^\infty dz\, \hat{a}_0 \Tr[F_{ij}F_{kz}]\epsilon_{ijk} \, . 
\eea
\end{subequations}
Implicitly, we have introduced the dimensionless gauge fields $a_i = \frac{A_i}{M_{\rm KK}}$, $a_u = A_U R(M_{\rm KK}R)^2$, 
and, as in the main text, $\hat{a}_0 = \frac{\hat{A}_0}{\lambda_0M_{\rm KK}}$. 

In the absence of instantons, the equation of motion for $x_4'(u)$ is given solely by $S_0$. With the boundary condition $x_4'(u_c)=\infty$ we find
\be \label{x42fix}
x_4'^2 = \frac{u_c^8f_T(u_c)}{u^3f_T(u)\left[u^8f_T(u)-u_c^8f_T(u_c)\right]} \, .
\ee
This is the leading order result for small instantons. (There are subleading contributions which we ignore, i.e., we work without backreactions of the instanton on the embedding of the 
flavor branes.) For small instanton widths,
 the integrands in $S_{\rm YM}$ and $S_{\rm CS}$ are nonzero only in a small vicinity around $z=\vec{x}=0$. 
This renders the abelian terms $\hat{F}_{0i}^2,\hat{F}_{0z}^2$ of higher order than the non-abelian ones $F_{ij}^2,F_{iz}^2$. Anticipating 
the eventual solution, one can do this systematically by rescaling 
$\vec{x} \to \vec{x}/\sqrt{\lambda}$, $z \to z/\sqrt{\lambda}$, and the gauge fields accordingly, and applying a systematic expansion for large $\lambda$. To keep the notation simple, we do not introduce rescaled quantities, 
but keep this expansion in mind, which yields leading and subleading contributions to the energy of the instanton that are eventually of first and zeroth order in $\lambda$, 
\be
S_{\rm YM}= S_{\rm YM}^{(1)} + S_{\rm YM}^{(0)} + \ldots
\ee
To compute these contributions, we first insert (\ref{x42fix}) into the YM action and change the integration variable from $u$ to $z$. Then, $S_{\rm YM}^{(1)}$ is obtained by an expansion around 
$z=0$ and dropping the abelian field strengths and all higher order terms ${\cal O}(z^2)$,
\be 
S_{\rm YM}^{(1)} =  \frac{\lambda N_cM_{\rm KK}}{96\pi^3T} \frac{u_c \sqrt{f_T(u_c)}}{\gamma}  \int d^3 x \int_{-\infty}^\infty dz\, \left(\frac{\Tr[F_{ij}^2]}{2}+\gamma^2\Tr[F_{iz}^2]\right) \, , 
 \ee
 with
 \be \label{gammaT}
 \gamma = \sqrt{6} u_c^{3/2}\sqrt{1-\frac{5u_T^3}{8u_c^3}} \, .
 \ee
Consequently, to leading order, the equations of motion for the non-abelian gauge fields yield the flat-space BPST solutions, with field strengths  
\be \label{FijFiz}
F_{ij}=\epsilon_{ija}\sigma_a\frac{2(\rho/\gamma)^2}{[x^2+(z/\gamma)^2+(\rho/\gamma)^2]^2} \, , \qquad F_{iz} = -\frac{\sigma_i}{\gamma}\frac{2(\rho/\gamma)^2}{[x^2+(z/\gamma)^2+(\rho/\gamma)^2]^2} \, ,
\ee
equivalent to the ansatz (\ref{ansatz1}) in the main text.

In order to compute the abelian field strengths, we go to subleading order, 
\bea 
S_{\rm YM}^{(0)} &=&  \frac{\lambda N_cM_{\rm KK}}{96\pi^3T}  \frac{u_c \sqrt{f_T(u_c)}}{\gamma} \int d^3 x \int_{-\infty}^\infty dz\, \left\{\lambda_0^2\frac{\Tr[\hat{F}_{0i}^2] + \gamma^2\Tr[\hat{F}_{0z}^2]}{f_T(u_c)}
+3u_cz^2\Tr[F_{iz}^2] \right.\non[2ex]
&&\left. +\frac{4u_c^6+10u_c^3u_T^3-5u_T^6}{8\gamma^2 u_c^5f_T(u_c)} z^2\left(\frac{\Tr[F_{ij}^2]}{2}+\gamma^2\Tr[F_{iz}^2]\right)\right\} \, . 
 \eea
The equation of motion for $\hat{a}_0$ becomes [recall that we work in Euclidean space, where $\hat{F}_{0i}^2 = -(\partial_i \hat{a}_0)^2$, $\hat{F}_{0z}^2 = -(\partial_z \hat{a}_0)^2$]
\be
\partial_i^2\hat{a}_0+\gamma^2\partial_z^2\hat{a}_0 = -\frac{3\gamma\sqrt{f_T(u_c)}}{4\lambda_0^2u_c}\Tr[F_{ij}F_{kz}]\epsilon_{ijk} \, ,
\ee
with the solution 
\be \label{a0xz}
\hat{a}_0(x,z) =  -\frac{3\sqrt{f_T(u_c)}}{2\lambda_0^2u_c}\frac{x^2+(z/\gamma)^2+2(\rho/\gamma)^2}{[x^2+(z/\gamma)^2+(\rho/\gamma)^2]^2} \, .
\ee
Inserting the solutions (\ref{FijFiz}) and (\ref{a0xz}) back into the action yields the 
energy  
\bea \label{E}
E\simeq T[S_{\rm YM}^{(1)} + S_{\rm YM}^{(0)}+S_{\rm CS}] = 18\pi^2\kappa M_{\rm KK} u_c\sqrt{f_T(u_c)}\left[1+\frac{9\gamma^2}{5\lambda_0^2u_c^2\rho^2}+\frac{u_c\beta\rho^2}{\gamma^2f_T(u_c)}\right] \, , \;\;
\eea
where the first term in the square brackets on the right-hand side comes from $S_{\rm YM}^{(1)}$, the second from the terms containing $\hat{a}_0$ in $S_{\rm YM}^{(0)}$ and 
$S_{\rm CS}$, and the third from the non-abelian contributions to $S_{\rm YM}^{(0)}$, and where we have abbreviated $\kappa = \frac{\lambda N_c}{216\pi^3}$ and
the temperature-dependent factor
\be
\beta \equiv 1-\frac{u_T^3}{8u_c^3}-\frac{5u_T^6}{16u_c^6} \, .
\ee  
Minimizing $E$ with respect to $\rho$ yields
\be \label{rhofull}
\rho^2 = \frac{12\pi}{\sqrt{5}\lambda}\frac{\gamma^2\sqrt{f_T(u_c)}}{u_c^{3/2}\beta^{1/2}} \, .
\ee
This solution justifies our expansion a posteriori since we now confirm that $S_{\rm YM}^{(1)} \sim {\cal O}(\lambda)$ and $S_{\rm YM}^{(0)},  S_{\rm CS}\sim {\cal O}(1)$.

\section{Equations of motion with symmetrized trace prescription}
\label{app:tseytlin}

Applying the symmetrized trace prescription (\ref{str}) to the DBI action (\ref{SDBI}) yields the Lagrangian  
\be 
{\cal L}= u^{5/2}\frac{(1+u^3f_T x_4'^2-\hat{a}_0'^2+2\bar{g}_1)(1+2\bar{g}_2)-\bar{g}_1\bar{g}_2}{\sqrt{(1+u^3f_T x_4'^2-\hat{a}_0'^2+\bar{g}_1)(1+\bar{g}_2)} } -n_I\hat{a}_0q(u) \, .
\ee
[to be compared to the Lagrangian from the unsymmetrized prescription (\ref{Lag})], where, for the sake of a compact notation in this appendix, we have abbreviated
\be
\bar{g}_1 = \frac{g_1}{3} \, , \qquad  \bar{g}_2 = \frac{g_2}{3} \, ,
\ee
with $g_1$, $g_2$ from Eqs.\ (\ref{g1g2sim}). 
The equations of motion in integrated form become
\begin{subequations}
\bea
\frac{u^{5/2}\hat{a}_0'[(1+2\bar{g}_2)(1+u^3f_Tx_4'^2-\hat{a}_0'^2)+\bar{g}_1\bar{g}_2]}{\sqrt{1+\bar{g}_2}(1+u^3f_Tx_4'^2-\hat{a}_0'^2+\bar{g}_1)^{3/2}} &=& n_I Q \, , \label{eom1a}\\[2ex]
\frac{u^{5/2}u^3f_Tx_4'[(1+2\bar{g}_2)(1+u^3f_Tx_4'^2-\hat{a}_0'^2)+\bar{g}_1\bar{g}_2]}{\sqrt{1+\bar{g}_2}(1+u^3f_Tx_4'^2-\hat{a}_0'^2+\bar{g}_1)^{3/2}} &=& k \, ,\label{eom2a}
\eea
\end{subequations}
with $Q$ as defined in the main part of the paper, Eq.\ (\ref{Qdef}). Dividing the first by the second equation yields  
\be
\frac{\hat{a}_0'}{u^3f_Tx_4'}=\frac{n_IQ}{k} \, , 
\ee
which can be used to write  
\be
1+u^3f_Tx_4'^2-\hat{a}_0'^2 = 1+\gamma_1 \hat{a}_0'^2 = 1+\gamma_2 x_4'^2 \, , 
\ee
with 
\be
\gamma_1 \equiv \frac{k^2}{u^3f_T\,(n_IQ)^2}-1 \, , \qquad \gamma_2 \equiv u^3f_T\left[1-\frac{u^3f_T\,(n_IQ)^2}{k^2}\right] \, .
\ee
This allows us to write both equations of motion in the form
\be \label{eomX}
X[(1+2\bar{g}_2)(1+X)+\bar{g}_1\bar{g}_2]^2=\eta(1+\bar{g}_2)(1+\bar{g}_1+X)^3 \, , 
\ee
where $X=\gamma_1\hat{a}_0'^2$ for Eq.\ (\ref{eom1a}) and $X=\gamma_2x_4'^2$ for Eq.\ (\ref{eom2a}),
and
\be
\eta\equiv \frac{k^2-u^3f_T(n_IQ)^2}{u^8f_T} \, .
\ee
The equations of motion (\ref{eomX}) can again be solved algebraically for $\hat{a}_0^2$, $x_4'^2$, as for the unsymmetrized prescription. However, now, they are cubic equations for $\hat{a}_0^2$, $x_4'^2$, which makes the solution
much more unwieldy. 

Next, we need to evaluate the stationarity equations for the free energy. For the minimization with respect to $n_I$, we have
\bea
\frac{\partial \Omega_{\rm baryon}}{\partial n_I} &=&\int_{u_c}^\infty du\left[ \frac{u^{5/2}}{2}\left(\frac{\partial \bar{g}_1}{\partial n_I}
\zeta_1+\frac{\partial \bar{g}_2}{\partial n_I}\zeta_2\right)+\hat{a}_0'Q\right] - \mu \, ,
\eea
where
\begin{subequations}
\bea
\zeta_1\equiv \frac{(1+u^3f_T x_4'^2-\hat{a}_0'^2)(3+4\bar{g}_2)+\bar{g}_1(2+3\bar{g}_2)}{(1+u^3f_T x_4'^2-\hat{a}_0'^2+\bar{g}_1)^{3/2}(1+\bar{g}_2)^{1/2}} \, ,  \\[2ex]
\zeta_2\equiv \frac{(1+u^3f_T x_4'^2-\hat{a}_0'^2)(3+2\bar{g}_2)+\bar{g}_1(4+3\bar{g}_2)}{(1+u^3f_T x_4'^2-\hat{a}_0'^2+\bar{g}_1)^{1/2}(1+\bar{g}_2)^{3/2}} \, .
\eea
\end{subequations}
A completely analogous calculation yields the derivatives with respect to $\rho$, $\gamma$, $z_0$, i.e., we have derived the analogues of Eqs.\ (\ref{mini1}) -- (\ref{mini4}), while the minimization with respect to 
$k$ is again given by Eq.\ (\ref{ell}). It remains 
to compute the derivative with respect to $u_c$, which is given by \cite{Li:2015uea},
\be
\frac{\partial\Omega_{\rm baryon}}{\partial u_c} = (k x_4' -{\cal L})_{u=u_c} +\int_{u_c}^\infty du\,\frac{\partial{\cal L}}{\partial u_c} \, ,
\ee
where the derivative in the second term is the {\it explicit} derivative with respect to $u_c$ (not acting on the $u_c$ dependence in $\hat{a}_0$, $\hat{a}_0'$, and $x_4'$).
For the first term, we need the following leading-order behaviors at $u\to u_c$,
\be
\bar{g}_1\simeq\frac{u_c f_T(u_c)\bar{\alpha}}{3\gamma_0(u-u_c)}  \, , \qquad \bar{g}_2\simeq \gamma_0\bar{\alpha}   \,, \qquad \eta \simeq  \frac{k^2}{u_c^8f_T(u_c)} \, ,
\ee
with $\gamma _0$ from Eq.\ (\ref{rhogamma}), $\bar{\alpha}\equiv \alpha/3$ with $\alpha$ defined in Eq.\ (\ref{alpha}), and 
\be
\gamma_2 \simeq  u_c^3f_T(u_c) \, , \qquad x_4' \simeq  \frac{c_1}{\sqrt{u-u_c}} \, , 
\ee
with $c_1$ given by the following cubic equation for $c_1^2$,
\be \label{C1}
3u_c^{10}\gamma_0^3f_T(u_c)c_1^2[3u_c^2c_1^2(1+2\gamma_0\bar{\alpha})+\bar{\alpha}^2]^2=k^2(1+\gamma_0\bar{\alpha})(3\gamma_0 u_c^2c_1^2+\bar{\alpha})^3 \, .
\ee
Using this equation, we compute
\be
(kx_4'-{\cal L})_{u=u_c} = -\frac{\bar{\alpha} k}{3u_c^2c_1\gamma_0\sqrt{u-u_c}}
\frac{3u_c^2c_1^2\gamma_0(3+4\bar{\alpha}\gamma_0)+\bar{\alpha}(2+3\gamma_0\bar{\alpha})}{3u_c^2c_1^2\gamma_0(1+2\gamma_0\bar{\alpha})+\bar{\alpha}^2\gamma_0}
+\frac{3\sqrt{3}\,\hat{a}_0(u_c) u_c^2\bar{\alpha}}{\sqrt{u-u_c}} \, , 
\ee
and 
\bea
\int_{u_c}^\infty du \, \frac{\partial {\cal L}}{\partial u_c} &=&\int_{u_c}^\infty du\left[ \frac{u^{5/2}}{2}\left(\frac{\partial \bar{g}_1}{\partial u_c}
\zeta_1+\frac{\partial \bar{g}_2}{\partial u_c}\zeta_2\right)+n_I\hat{a}_0'\frac{\partial Q}{\partial u_c}\right] -  \frac{3\sqrt{3}\,\hat{a}_0(u_c) u_c^2\bar{\alpha}}{\sqrt{u-u_c}} \, . \;\;
\eea
Consequently, the minimization with respect to $u_c$ becomes
\bea
0&=&\int_{u_c}^\infty du \,\left[\frac{u^{5/2}}{2}(\bar{g}_1\zeta_1p_-+\bar{g}_2\zeta_2p_+)+n_I\hat{a}_0'\frac{\partial Q}{\partial u_c} \right.\non[2ex]
&&\left.-\frac{\bar{\alpha} k (u-u_c)^{-3/2}}{6u_c^2\gamma_0c_1}
\frac{3u_c^2c_1^2\gamma_0(3+4\bar{\alpha}\gamma_0)+\bar{\alpha}(2+3\gamma_0\bar{\alpha})}{3u_c^2c_1^2\gamma_0(1+2\gamma_0\bar{\alpha})+\bar{\alpha}^2\gamma_0}
+\frac{3u_c^2}{u^{1/2}f_c}\frac{\bar{g}_1\zeta_1}{2}\right] \, .
\eea
This completes the set of equations needed to find the ground state for the case of the symmetrized trace prescription. The equations are very similar to the simpler ones of the 
unsymmetrized case. The main difference are the cubic equations whose solutions enter the integrands of all stationarity equations, making the numerical evaluation more involved.

\section{Calculation of phase transition lines and critical endpoints}
\label{app:phases}

In this appendix we explain our calculation of the various phase transition lines for the phase diagrams in Fig.\ \ref{fig:phases}. It is obviously very tedious to compute the ground state on a grid in the 
$\rho_0$-$\gamma_0$-$\mu$ space and deduce the phase transition lines from this calculation. More efficiently, after getting an idea of the overall structure of the phase diagram, one may proceed as follows. 

\begin{itemize}

\item {\it Chiral phase transition.} This is the first-order phase transition between the chirally symmetric phase and the baryonic phase, i.e., the phase transition line is defined by 
$\Omega_{\rm baryon} = \Omega_{\rm quark}$. We solve this equation simultaneously with Eqs.\ (\ref{mini1}), (\ref{mini4}), and (\ref{mini5}) for the variables $k$, $n_I$, $z_0$, and $\mu$ [all rescaled with appropriate 
powers of $u_c$, such that Eq.\ (\ref{ell}) decouples]. The baryonic phase at this transition can have either $z_0=0$ ($N_z=1$) or $z_0>0$ ($N_z=2$). The transition with $z_0=0$ can be computed separately without the minimization with respect to $z_0$ (\ref{mini4}), and the intersection of the two phase transition lines defines a critical point, see upper left corner of both panels in Fig.\ \ref{fig:phases}.  

\item {\it Onset of second instanton layer.} This is the transition within the baryonic phase which separates $z_0=0$ from $z_0>0$ ($z_0>0$ only appears for $N_z=2$, no higher number of instanton layers is preferred). 
This transition can be either second or first order. The second order transition line 
is determined as explained in the context of Fig.\ \ref{fig:absence}: after dividing the minimization with respect to $z_0$ (\ref{mini4}) by $z_0$, we take the limit $z_0\to 0$, i.e., we approach the transition form the $z_0>0$ side. The resulting equation
is solved simultaneously with Eq.\ (\ref{mini5}) in the limit $z_0\to 0$ for the variables $n_I$ and $k$, which are then used to compute the critical chemical potential with the help of Eq.\ (\ref{mini1}). The calculation 
of the first-order onset of the second layer is a little more complicated because we have to compare the free energies of the phases with one and two layers.
We need to solve the following 7 equations simultaneously:
Eqs.\ (\ref{mini1}) and (\ref{mini5}) for $z_0=0$, Eqs.\ (\ref{mini1}), (\ref{mini4}), and (\ref{mini5}) for $z_0>0$, plus the two conditions that the chemical potentials and the free energies of both phases are identical. The corresponding 7 variables are $k$, $n_I$, $\mu$ for both phases and $z_0$ for the $z_0>0$ phase (as always we work with quantities rescaled by $u_c$, so, more precisely, the two chemical potentials used as variables in our calculation are $\tilde{\mu}_{1} = \mu_{1}/u_{c,1}$, $\tilde{\mu}_{2} = \mu_{2}/u_{c,2}$, and at the phase transition $\mu_{1} = \mu_{2}$, while 
in general $\tilde{\mu}_{1}\neq \tilde{\mu}_{2}$). The critical point that separates the second-order from the 
first-order transition line can be found by asking at which $\gamma_0$ the curve $z_0(\mu)$ becomes multivalued, which is equivalent to asking at which $\gamma_0$ this curve bends to the left at $z_0\to 0$. 

\item {\it Baryon onset.} This is the phase transition between the mesonic and baryonic phases. Therefore, it is defined by $\Omega_{\rm baryon} = \Omega_{\rm meson}$, and this equation has to be solved simultaneously 
with Eqs.\ (\ref{mini1}), (\ref{mini4}), and (\ref{mini5}) for the variables $k$, $n_I$, $z_0$, and $\mu$. This is completely analogous to the chiral phase transition. Again, we need to compute the two cases $z_0=0$ and $z_0>0$.
The former transition can be either first order or second order, and the second order transition can be found by solving the single equation (\ref{mini5}) for $k$ with an infinitesimally small $n_I\to 0$.

\end{itemize}
\bibliographystyle{JHEP}
\bibliography{references}

\end{document}